\documentclass{aa}  

\usepackage[dvipsnames]{xcolor}
\usepackage{graphicx}
\usepackage{txfonts}
\usepackage{multirow}
\usepackage{multicol}
\usepackage{makecell}
\usepackage{caption}
\usepackage{float}
\usepackage{verbatim}
\DeclareCaptionFormat{cont}{#1 (cont.)#2#3\par}
\usepackage{esvect}
\usepackage[multiple]{footmisc}  
\usepackage{supertabular,booktabs}
\usepackage{afterpage}
\usepackage{CJK}
\usepackage[flushleft]{threeparttable}
\usepackage{xfrac}
\usepackage{subfig}
\usepackage{soul}
\usepackage{cancel}

\begin{document} 
\title{Pulsar Scintillation Studies with LOFAR: \uppercase\expandafter{\romannumeral1} The Census}
\author{
  Ziwei Wu \inst{1} 
  \and 
   Joris P.W. Verbiest \inst{1, 2} 
   \and
   Robert A. Main \inst{2} 
   \and
   Jean-Mathias Grie{\ss}meier \inst{3,4} 
   \and
   Yulan Liu \inst{1} 
   \and
   Stefan Os\l owski \inst{5} 
   \and
   Krishnakumar Moochickal Ambalappat \inst{2, 1} 
   \and
   Ann-Sofie Bak Nielsen \inst{2,1} 
   \and
   J\"orn K\"unsemöller \inst{1} 
   \and
   Julian Y. Donner \inst{2,1} 
    \and
   Caterina Tiburzi \inst{6} 
    \and
   Nataliya Porayko \inst{2} 
    \and
   Maciej Serylak \inst{7,8} 
   \and
   Lars K\"unkel\inst{1} 
   \and
   Marcus Brüggen\inst{9} 
   \and
   Christian Vocks\inst{10} 
}

   \institute{
        Fakult\"at f\"ur Physik, Universit\"at Bielefeld, Postfach 100131, 33501 Bielefeld, Germany \\
        \email{zwu@physik.uni-bielefeld.de}
        \and
        Max-Planck-Institut f\"ur Radioastronomie, Auf dem H\"ugel 69, 53121 Bonn, Germany
        \and
        LPC2E - Universit\'{e} d'Orl\'{e}ans / CNRS, 45045071 d'Orl\'{e}ans cedex 2, France 
        \and
        Station de Radioastronomie de Nan\c{c}ay, Observatoire de Paris - CNRS/INSU, USR 
        704 - Univ. Orl\'{e}ans, OSUC, Route de Souesmes, 18330 Nan\c{c}ay, France
        \and
        Manly Astrophysics, 15/41-42 East Esplanade, Manly, NSW 2095, 
	    Australia 
        \and
        ASTRON, the Netherlands Institute for Radio Astronomy, Oude Hoogeveensedijk 4, Dwingeloo 7991 PD, the Netherlands
        \and
        SKA Observatory, Jodrell Bank, Lower Withington, Macclesfield, SK11 9FT, United Kingdom
        \and
        Department of Physics and Astronomy, University of the Western Cape, Bellville, Cape Town 7535, South Africa
        \and
        Hamburger Sternwarte, University of Hamburg, Gojenbergsweg 112, 21029 Hamburg, Germany
        \and
        Leibniz-Institut f\"ur Astrophysik Potsdam (AIP), An der Sternwarte 16, 14482 Potsdam, Germany
             }
\date{Received **; accepted **}

\abstract
    {Interstellar scintillation (ISS) of pulsar emission can be
      used both as a probe of the ionised interstellar medium (IISM) and
      cause corruptions in pulsar timing experiments. 
      Of particular interest are so-called scintillation arcs which
      can be used to measure
      time-variable interstellar scattering delays directly,
      potentially allowing
      high-precision improvements to timing precision.}
    {The primary aim of this study is to carry out the first sizeable
      and self-consistent census of diffractive pulsar scintillation and
      scintillation-arc detectability at low frequencies, as a primer
      for larger-scale IISM studies and pulsar-timing related
      propagation studies with the LOw-Frequency ARray (LOFAR) High
      Band Antennae (HBA).}
    {We use observations
      from five international LOFAR stations and
      the LOFAR core in
      the Netherlands. We analyse the 2D auto-covariance function of
      the dynamic spectra of these observations to determine the
      characteristic bandwidth and time-scale of the ISS towards the
      pulsars in our sample and investigate the 2D power spectra of the dynamic spectra to determine the presence of scintillation arcs.
    }
    {
      In this initial set of 31 sources, 15 allow
      full determination of the scintillation properties; nine of
      these show detectable scintillation arcs at 120-180 MHz.
      Eight of the observed sources show unresolved scintillation;
      and the final
      eight don't display diffractive scintillation. Some correlation
      between scintillation detectability and pulsar brightness and
      dispersion measure is apparent, although no clear cut-off values
      can be determined. 
      Our measurements across a large fractional bandwidth allow a
      meaningful test of the
      frequency scaling of scintillation parameters, uncorrupted by
      influences from refractive scintillation variations.
    }
    {Our results indicate the powerful advantage and great
      potential of ISS studies at low frequencies and
      the complex dependence of scintillation detectability on
      parameters like pulsar brightness and interstellar
      dispersion. This work provides the first installment of a larger-scale
      census and longer-term monitoring of interstellar scintillation
      effects at low frequencies.
    }


    \maketitle
    %

    \section{Introduction}
    Radio pulsars are neutron stars that emit beams of radio waves from
    their magnetic poles. Due to the small size of these stars and the
    even smaller size of their emission regions, pulsars are
    detectable as point sources. The compact beams of radiation
    emitted from the pulsar's magnetic poles are constantly perturbed
    by refractive index fluctuation in the ionized interstellar medium
    (IISM), generating random phase variations in the various rays of
    light.  The interference between these essentially uncorrelated
    scattered rays 
    results in a modulation of the pulse intensity as a function of frequency
    and time, 
    which is well-known as
    interstellar scintillation \citep[ISS,][]{sch68}. The two main
    types of ISS are diffractive ISS \citep[DISS,][]{ric69} caused by
    small-spatial-scale density fluctuations ($10^{6}-10^{8}$ m) and
    refractive ISS (RISS, \citealt{sie82, rcb84}) resulting from
    large-spatial-scale density inhomogeneities ($10^{10}-10^{12}$ m)
    in the IISM. These types become distinct in the strong scattering
    regime (multi-path propagation) where numerous scattered rays
    interfere with one another to form an interference pattern on the
    observer plane \citep{r90, n92}.  However, the origin of ISS
    is still under debate and some alternative models based on
    discrete plasma structures have been proposed recently
    \citep{rbc87,pl14,gwi19}. 

    Scintles, i.e.\ enhanced pulse intensity variations with relatively
    short time scales and narrow frequency bandwidths, are identified in
    dynamic spectra, which are a
    two-dimensional matrix of pulsed intensity as a function of
    time ($t$) and frequency ($\nu$) (see Figure.~\ref{fig:scintillation_census}).  
    With diffractive interstellar scintillation parameters
    obtained from dynamic spectra, one can study the turbulence in the IISM \citep{cwb85, sg90}, the local bubble
    \citep{bgr98}, constrain the pulsar proper motion
    \citep{c86, g95}, study properties of binary systems \citep{cmr+05,
      rcn+14}, modulations on DISS from RISS \citep{brg99b}, model the
    IISM based on annual variations of scintillation
    \citep{rch+19}, etc.

    Two decades ago, scintillation arcs were identified in the
    so-called 
    secondary spectra, which are the 2D power spectra of dynamic
    spectra \citep{smc+01}. These scintillation arcs probe the IISM
    structure and frequently show up as `criss-cross' sloping
    bands in dynamic spectra, resulting from interference between rays
    in a central core and scattered rays from an extended scattering
    disc \citep{wms+04, crs+06}. The past decade-and-a-half have shown
    increasing interest and applications of these arcs, see e.g.\
    \citet{tr07,wks+08,bmg+10,pl14, msa+20, rcb+20, yzm+21}.

    High-precision pulsar timing experiments, such as pulsar timing
    arrays (PTAs), are a promising method of detecting and
    characterizing low-frequency gravitational waves \citep[see,
      e.g.\ ][and references therein]{vob21}.  PTA experiments
    currently rely on stable millisecond pulsars with low dispersion
    measure (DM\footnote{This parameter quantifies
      the integrated electron density between us and the pulsar: $DM =
      \int_0^D n_{\rm e}{\rm d}l$.} < 50 $\rm{pc\,cm^{-3}}$), at high frequencies (mostly at 1.4 GHz) to minimize
    propagation effects on pulsar timing precision \citep{lcc+17}.
    The two branches of propagation effects that could affect the
    pulsar timing precision are dispersion and scattering
    \citep{vs18}.  Dispersion is well-studied and its effect on pulsar
    timing can in principle be measured precisely and eliminated
    completely
    \citep{dvt+20}; although, see \cite{css16} for a complicating factor.
    Scintillation and the related
    pulse broadening \citep{cr98}, however, are less easily corrected
    \citep{lkd+17}.
    Moreover, at high observing frequencies, nearby pulsars could be in the
    weak scintillation regime, resulting in a small number of scintles
    in the dynamic spectrum and with relatively small pulse broadening
    delays. A possible way to mitigate the propagation effects on PTA
    data at high frequencies could be through low-frequency monitoring
    of these pulsars, to determine corrections for the high-frequency
    data. The scattering time delays could then potentially be
    measured directly from the power
    distribution in the secondary spectrum \citep{hs08, msa+20} or through
    holographic techniques \citep{wks+08,pmd+14, dem11, wdv13}, which may lead to
    significant improvements in timing precision. 

    In this work, we present the first census of scintillating pulsars
    with the LOw Frequency ARray (LOFAR). 
    This work has been organized in the
    following manner: in Section 2 we describe our observations and
    data processing; in Section 3 we show the analysis and results.
    Section 4 contains our conclusions.

    \section{Observations and data processing}
    \subsection{Observations}
    Our analysis is based on observations in the frequency range
    120-180 MHz, taken in stand-alone mode with five international LOFAR
    stations \citep{vwg+13}, namely those in Effelsberg
    (DE601), Tautenburg (DE603), Potsdam-Bornim (DE604),
    Norderstedt (DE609) and Nan\c{c}ay (FR606),
    as well as the LOFAR core (see Table~\ref{tab:properties} for
    details).
    Our processing pipeline was based on the \textsc{DSPSR}
    \citep{vb11} package with frequency and time resolution tuned
    depending on the scintle size. 
    Subsequently, observations were
    written out in the \textsc{PSRFITS} format \citep{hvm04} and processed with
    \textsc{PSRCHIVE} \citep{vdo12}.

    \subsection{Source selection}
    
\begin{figure}
    \centering
    \caption{Source sample selection. Shown are all pulsars with DM
      lower than 50\,pc/cm$^3$. The star symbols represent pulsars
      with a published flux density at 150 MHz. The dot markers
      represent the pulsars for which no published flux density at 150
      MHz exists; in these cases the flux densities were extrapolated
      from higher frequencies, using a conservative spectral index of
      $-1.4$ (unless a spectral-index measurement was previously
      published). The green symbols represent pulsars for which we
      detected scintillation in our LOFAR data, while red symbols
      represent pulsars for which scintillation was not
      detectable. Orange symbols indicate sources for which the
      scintillation was detected but not resolved (see
      text). Sources for which scintillation arcs were published in
      literature are encircled with a grey circle, while scintillation
      arcs presented in this work are represented by green circles.}
    \includegraphics[width=1.0\linewidth]{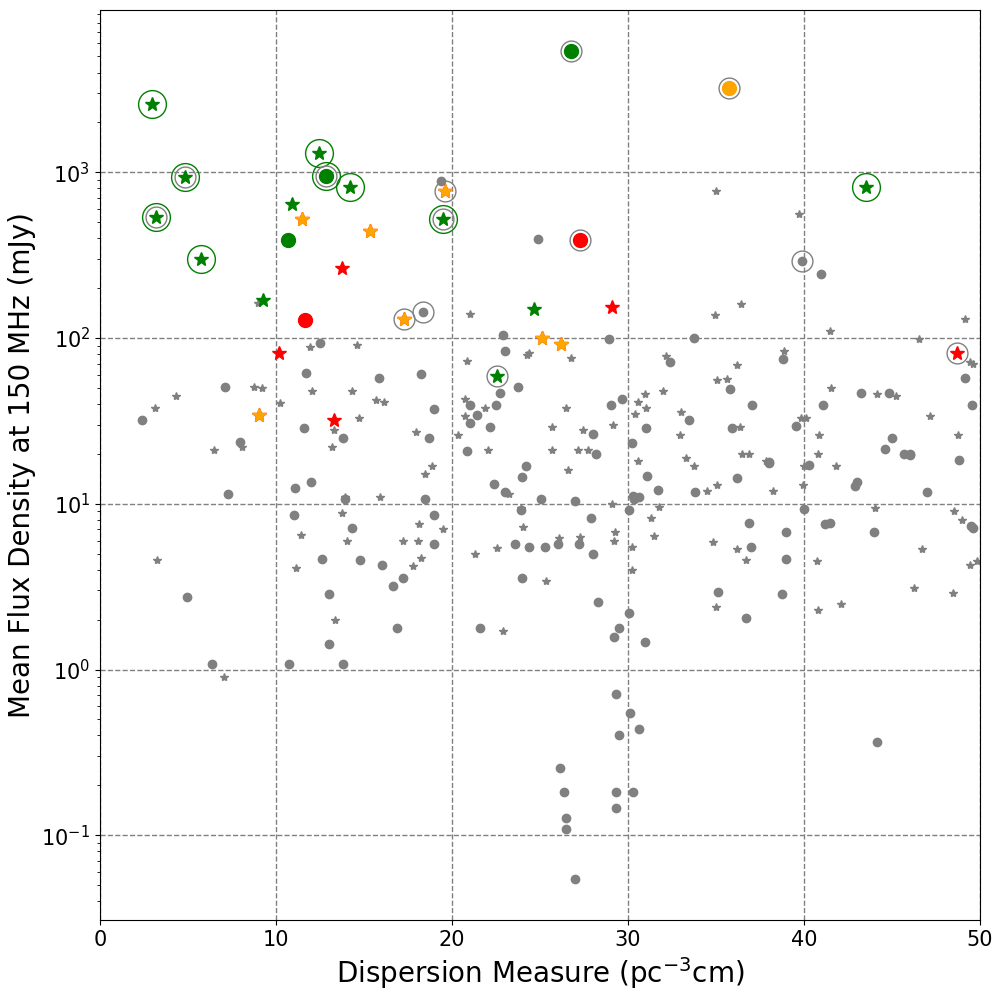}
    \label{fig:sample}
    \end{figure}

    The pulsars included in this work were selected primarily based on
    their dispersion measure, and on their brightness in the
    LOFAR HBA band, as given by \citet{kvh+16}, \citet{bkk+16},
    \citet{scb+19} or \citet{xbt+17}, or based on extrapolations from
    higher frequencies.
    Based on the thin-screen theory, the scintillation
    bandwidth is expected to scale as $\Delta \nu_{\rm d} \propto
    DM^{-2.2} \nu^{4.4}$ \citep[e.g.,][]{rnb86}, so that large values for DM would result in
    unresolvably small scintles at our low frequencies. 
    Consequently, we only considered pulsars in the LOFAR sky
    (declination above $-20^{\circ}$) with a DM below 50 pc/cm$^3$ and
    with a flux density at 150\,MHz above 10\,mJy. To evaluate the
    flux density we either used published flux densities at 150\,MHz
    directly, or extrapolated from higher frequencies based on the
    known spectral index of the pulsar in question or based on a
    spectral index of $-1.4$, which is slightly shallower than the
    expected average spectral index of radio pulsars
    \citep{blv13,jvk+18} and hence provides a slightly conservative
    estimate of the flux density at 150\,MHz. Pulsars that were given
    as non-detections by either \citet{kvh+16} or \citet{bkk+16} were
    excluded. All pulsars that satisfy the declination and DM
    requirements and for which an estimate of the flux density could
    be obtained, are shown in
    Figure~\ref{fig:sample}. The brightness of the pulsar affects both the ability to
    detect scintillation arcs (because only a small fraction of the
    emitted power will be seen spread out to the arcs) and to detect
    scintles (since the high resolution required for scintillation
    studies at these frequencies implies the S/N threshold for
    detection must be achievable during very short durations and
    across very narrow channels). However, since no previous clearly
    defined census of scintillation was carried out, particularly at
    low frequencies, no clear cut-off value for the flux density can
    be defined. Consequently in the selection of this initial
    installment of the census, we concentrated on bright pulsars,
    while also including some fainter sources that appeared promising,
    based on scintillation studies at higher frequencies. 

    \subsection{Data processing}

    \subsubsection{Radio-frequency interference (RFI)}

    The RFI cleaning program
    \textsc{iterative\_cleaner}\footnote{Available from
      \url{https://github.com/larskuenkel/iterative_cleaner}} is a
    modification of the \textsc{surgical} method included in the RFI
    cleaner of the \textsc{COASTGUARD} pulsar-data analysis package
    \citep{lkg+16}. Two major changes were made: first, the
    \textsc{iterative\_cleaner} uses an iterative approach to
    determine the RFI-free template
    profile, which is particularly useful when the RFI is more
    powerful than the pulsar signal. Second, the (simple) de-trending algorithm for
    correcting gradual changes in both time and frequency was removed.

    \subsubsection{Scintillation parameters}

    After polarisation averaging to total intensity using the
    \textsc{pam} program of the \textsc{PSRCHIVE} package, we created
    the initial dynamic spectrum with the \textsc{dynamic\_spectra} or
    \textsc{psrflux} programs (also from the \textsc{PSRCHIVE}
    package).  Trends in the frequency direction as a result of the
    instrumental bandpass and the spectral index of the pulsar are
    then removed from the dynamic spectrum by fitting a power-law
    function to a time-averaged version of the dynamic spectrum.  
    In order to minimize the impact of temporal variations on the dynamic spectrum, we subtract the mean of each sub-integration. This approach was previously introduced by \cite{rch+19}.

    Using the dynamic spectrum, one can estimate the diffractive
    scintillation bandwidth $\Delta\nu_{\rm{d}}$ and the diffractive
    scintillation time-scale $\tau_{\rm{d}}$ by computing a
    two-dimensional autocovariance function (2D ACF) of the dynamic
    spectrum. To calculate the 2D ACF, we first pad the finite dynamic
    spectrum with an equal length of zeroes in the frequency and time
    dimensions, then perform a 2D fast Fourier transform (FFT), take
    the squared magnitude of the result, and perform an inverse 2D
    FFT, following \cite{rch+19}.  Since the center of the 2D ACF is
    often visible as a noise bridge, it is replaced by the mean value
    of nearby pixels.  Finally the 2D ACF is peak-normalized.

    Next, we determine the scintillation bandwidth and timescale by
    fitting a Gaussian and an exponential function to one-dimensional cuts through the
    2D ACF along the X and Y axes ($\nu = 0$ and $\tau = 0$,
    respectively). This approach ignores any tilts in the ACF, but
    suffices as a first-pass analysis of the ISS properties along our
    lines of sight to pulsars. The analysis of tilts in our ACFs is deferred to a
    future paper. The functions fitted are (a slight modification to the previous standard of a Gaussian function, \citealt{cr98}):
    \begin{equation}
      \begin{split}
        \rm{ACF}(\nu = 0, \tau) &= \rm{exp}(-a*\tau^{2}) \\ 
        \rm{ACF}(\nu, \tau = 0) &= \rm{exp}(-b*\nu) 
	    \label{eq:fit}
      \end{split}
    \end{equation}
    and the resulting scintillation parameters are defined as (similar to \citealt{brg99}): 
    \begin{equation}
      \begin{split}
        \tau_{d,obs} &= \sqrt{\frac{1}{a}} \\
        \Delta \nu_{d,obs} &= \rm{\frac{\ln{2}}{b}}.
	    \label{eq:fit_result}
      \end{split}
    \end{equation}
\begin{table*}[ht]
\begin{threeparttable}
\centering
\caption{Properties of the observations and scintillation
  characteristics for the 15 pulsars with detectable scintillation. 
}
\begin{tabular}{lrcrcccrllrrl}
\toprule \toprule
PSR Name & DM & Period & Station & Date & Length& $\Delta f$ & $\Delta t$ & \multicolumn{1}{c}{$\Delta \nu_{d}^{a}$} & \multicolumn{1}{c}{$\tau_{d}^{a}$} & \multicolumn{1}{c}{$u$} & \multicolumn{1}{c}{$t_{\rm{r}}$} & \multicolumn{1}{c}{$\alpha$} \\ 
(J2000) & (pc/cm$^3$) & (s) & & & (hours) & (kHz) & (s) & \multicolumn{1}{c}{(kHz)} & \multicolumn{1}{c}{(min)} &  & \multicolumn{1}{c}{(Day)} & \\ 
\midrule
J0034$-$0721 & 10.9 & 0.943 & FR606 & 2020-10-01 & 1.0  & 5   & 10 & \phantom{0}67.1(5) & \multicolumn{1}{c}{--} &67  & -- & 5.0(5)\\
J0332+5434   & 26.8 & 0.715 & FR606 & 2020-12-08 & 0.5  & 0.3 & 10 & \phantom{0}\phantom{0}0.91(3) & \phantom{0}0.78(0) &570 & 206 & 4.46(6)\\
J0814+7429   &  5.8 & 1.292 & DE604 & 2017-04-29 & 3.0  & 5   & 10 & 326(16) & 19.3(9) &30  & 13& 4.4(7)\\
J0826+2637   & 19.5 & 0.531 & Core  & 2019-12-03 & 0.5  & 1   &  5 & \phantom{0}\phantom{0}5.28(2) & \phantom{0}0.64(0) &240 & 26   & 4.1(2)\\
J0837+0610   & 12.9 & 1.274 & DE601 & 2020-01-19 & 2.0  & 5   & 10 & \phantom{0}\phantom{0}9.95(4) & \phantom{0}1.79(1) &174 & 40   & 4.5(1)\\
J0953+0755   &  3.0 & 0.253 & DE601 & 2016-01-04 & 5.0  &195  & 60 & 916(57) & 18(1)   &18  & 4    & 3.7(6)\\
J1136+1551   &  4.8 & 1.188 & DE601 & 2015-04-10 & 2.0  & 5   & 10 & \phantom{0}\phantom{0}6.3(2)  & \phantom{0}0.48(0) &218 & 13   & 4.1(5)\\
J1239+2453   &  9.3 & 1.382 & FR606 & 2020-05-20 & 1.0  & 5   & 10 & \phantom{0}36.6(4) & \phantom{0}2.15(2) &91  & 14   &  3.2(3)\\
J1607$-$0032 & 10.7 & 0.422 & FR606 & 2020-09-08 & 1.0  & 1.25& 10 & \phantom{0}20.8(6) & \phantom{0}6.8(1)  &120 & 137 & 4.6(3)\\
J1921+2153   & 12.4 & 1.337 & DE609 & 2018-08-26 & 1.7  & 5   & 10 & \phantom{0}22.1(1) & \phantom{0}1.1(0)  &116 & 11    & 4.0(3)\\
J1932+1059   &  3.2 & 0.227 & FR606 & 2020-12-30 & 1.0  & 5   & 10 & \phantom{0}59.8(9) & \phantom{0}3.2(1)  &71  & 13   & 4.0(4)\\
J2018+2839   & 14.2 & 0.558 & DE603 & 2019-12-27 & 2.0  & 5   & 10 & \multicolumn{1}{c}{--} & \phantom{0}4.51(4) & --  & --    & -- \\
             &      &       & FR606 & 2020-12-15 & 1.0  & 0.3 & 10 & \phantom{0}\phantom{0}2.9(8)  & \phantom{0}3.95(2) & 321& 632  & 4.2(3)\\ 
J2022+2854   & 24.6 & 0.343 & Core  & 2019-12-03 & 0.5  & 1   &  5 & \phantom{0}\phantom{0}5.88(4) & \phantom{0}2.05(1) & 226& 97   & 6.6(3)\\
J2022+5154   & 22.6 & 0.529 & Core  & 2019-12-03 & 0.5  & 1   &  5 & \phantom{0}\phantom{0}4(5)    & \multicolumn{1}{c}{--} & 280 & --   & -- \\  
J2219+4754   & 43.5 & 0.538 & Core  & 2020-01-15 & 0.5  & 0.08&  5 & \phantom{0}\phantom{0}0.23(1) & \phantom{0}0.54(0) & 1142&572  & 4.3(2)\\ \bottomrule
\end{tabular} 
\begin{tablenotes}
\small
\item {\bf Notes:} Given are the pulsar name, dispersion measure and period
  along with the LOFAR station used, the date and length of the observation, the
  frequency and time resolution used and the measured scintillation
  parameters. $u$ is the scattering strength (see
  Equation~\ref{eq:scintillation strength}) and $t_{\rm r}$ is the
  estimated refractive timescale at LOFAR frequencies (see
  Equation~\ref{eq:ref_sci_time_scale}). Numbers in brackets denote the formal 1-$\sigma$ uncertainty in the
  last digit quoted.
\item [a] The given scintillation bandwidths $\Delta \nu_{d}$ and scintillation
  time-scales $\tau_{d}$ are for a centre frequency of 150\,MHz and
  have been measured over the range 120-180\,MHz for all but three
  pulsars: for PSRs~J0332+5434 and J2018+2839 the frequency range
  140-160\,MHz was used; and for PSR~J2219+4757 the range 145-155\,MHz
  was used. 
\end{tablenotes}
\label{tab:properties}
\end{threeparttable}
\end{table*}
    
\begin{table*}[ht]
\begin{threeparttable}
\centering
\caption{Observational data for pulsars for which the scintillation bandwidth $\Delta \nu_{d}$
  could not be
  successfully derived. }
\begin{tabular}{lrcrcccccc}
\toprule \toprule
PSR Name & DM & Period & Station & Date & Length& $\Delta f$ & $\Delta t$  & Ref & Note$^a$\\ 
(J2000) & (pc/cm$^3$) & (s) & & & (hours) & (kHz) & (s) & & \\
\midrule
J0034$-$0534 & 13.8 & 0.002   & FR606 & 2020-12-29 & 1.0 & 1.25& 10 & &\\
J0323+3944 & 26.2 & 3.032   & FR606 & 2021-03-17 & 1.0 & 0.3 & 10 &
\citet{sw85} & low S/N\\
J0922+0638 & 27.3 & 0.431   & FR606 & 2020-09-09 & 1.0 & 1.25& 10 & \citet{brg99} &\\
J0946+0951 & 15.3 & 1.098   & FR606 & 2020-12-30 & 1.0 & 1.25& 10 & &low S/N \\
J1012+5307 & 9.0  & 0.005   & Core  & 2020-12-03 & 0.5 & 32  & 5  & \citet{lmj+16} & low S/N\\
J1300+1240 & 10.2 & 0.006   & FR606 & 2021-03-18 & 1.0 & 0.3 & 10 & \citet{gg00} &\\
J1509+5531 & 19.6 & 0.740   & FR606 & 2020-12-01 & 1.0 & 0.3 & 10 &
\citet{brg99} & low S/N\\
J1537+1155 & 11.6 & 0.038   & FR606 & 2020-12-30 & 1.0 & 1.25& 10 & \citet{jnk98} &\\
J1645$-$0317 & 35.8 & 0.388   & DE601 & 2021-11-30 & 2.0 & 5   & 10 & 
\citet{sss+06} & insufficient $\Delta f$ \\
J1740+1311 & 48.7 & 0.803   & FR606 & 2020-06-25 & 1.0 & 0.16& 10 & \citet{crs+06}&\\
J1857+0943 & 13.3 & 0.005   & FR606 & 2021-03-18 & 1.0 & 0.6 & 10 & \citet{lmj+16} &\\
J1959+2048 & 29.1 & 0.002   & Core  & 2021-07-26 & 0.5 & 32  & 10 & \citet{mvp+17} &\\
J2048$-$1616 & 11.5 & 1.962   & FR606 & 2020-09-08 & 1.0 & 1.25& 10 &
\citet{brg99} & low S/N\\
J2113+2754 & 25.1 & 1.203   & FR606 & 2021-03-18 & 1.0 & 0.6 & 10 &   & low S/N \\
J2313+4253 & 17.3 & 0.349   & FR606 & 2021-03-17 & 1.0 & 0.3 & 10 &
\citet{brg99} & low S/N\\
J2330$-$2005 & 8.5  & 1.644   & FR606 & 2020-09-08 & 1.0 & 1.25& 10 &
\citet{brg99} & \\
\bottomrule
\end{tabular}
\begin{tablenotes}
\small
\item {\bf Notes:} Columns as in Table~\ref{tab:properties},
  except for the final two columns, which give the reference to
  earlier published scintillation results for some sources and a note
  on the likely cause for our non-detections. For pulsars that have
  published scintillation results at other frequencies, the relevant
  reference is given.
\item [a] ``low S/N'' implies that scintles were detected, but with
  insufficient S/N to allow robust measurements of the scintillation
  statistics; in all other cases likely a combination of lacking
  sensitivity and frequency resolution would have contributed to the
  absence of scintles in our data. "insufficient $\Delta f$" means that the used frequency resolution is insufficient to resolve the scintillation to get the meaningful measurements.
\end{tablenotes}
\label{tab:more_tested_pulsars}
\end{threeparttable}
\end{table*}
    
The uncertainties of the individual points in the one-dimensional
    ACF slices are given by 
    Eq.~1 of
    \citet{brg99}. Due to the limited time and frequency resolution
    of our dynamic spectra, the scintles appear bigger than they
    really are. This effect in both these parameters can be corrected
    for by subtracting the resolution from the parameters
    quadratically, as described by \citet{brg99}:
    \begin{equation}
      \begin{split}
        \Delta\nu_{d} &= \sqrt{\Delta \nu_{d,obs}^{2} - \Delta f^{2}} \\
        \tau_{d} &= \sqrt{\tau_{d,obs}^{2}-\Delta t^{2}},
      \end{split}
    \end{equation}
    where $\Delta f$ and $\Delta t$ are the frequency and time
    resolution (listed in Table~\ref{tab:properties}), respectively. The
    uncertainty of the scintillation parameters consists of the
    quadrature sum of the
    uncertainty coming from the fitting procedure and the statistical
    error $\sigma_{\rm{est}}$ due to the finite number of scintles
    \citep{brg99}: 
    \begin{equation}
      \rm{ \sigma_{est} = (f_{d}*\frac{BW_{\rm{dyn}}T_{\rm{dyn}}}{\Delta\nu_{\rm{d}} \tau_{\rm{d}}})^{-0.5}.}
	  \label{eq:error}
    \end{equation}
    Here $\rm{BW_{dyn}}$ and $\rm{T_{dyn}}$ are the observing bandwidth
    and length, respectively, and $f_{d}$ (= 0.4)
    is the filling factor. For PSR~J0953+0755, $\sigma_{\rm{est}}$ is
    about 17\%. In contrast, for all other pulsars in our sample,
    $\sigma_{\rm{est}}$ is typically
    smaller than 1\% in the LOFAR frequency range, even
    in 10-MHz-wide sub-bands (see Section~\ref{sec:results}).

    To compute the secondary spectra, following \cite{rcb+20}, we apply a Hamming window
    function to the outer 10\% of each dynamic spectrum to reduce the effects of aliasing in the secondary spectrum. After
    this, we form the secondary spectrum using a 2D discrete Fourier
    transform, taking its squared magnitude, shifting it, 
    and then
    converting the relative power levels into a decibel scale.  
    The program \textsc{parabfit} described in \cite{bot+16} is used to measure the arc curvature based on a Hough transform. Fundamentally, \textsc{parabfit} sums the power over a given parabolic region in the secondary spectrum; and optimises the width (parametrised by the \textsc{pdist} parameter) and opening angle (parameterised by the \textsc{curves} parameter) of the parabolae in order to achieve a maximum of summed power.
    
    \section{Analysis and results}\label{sec:results}
\begin{figure*}[hbtp]
     \centering
     \includegraphics[width=0.195\linewidth]{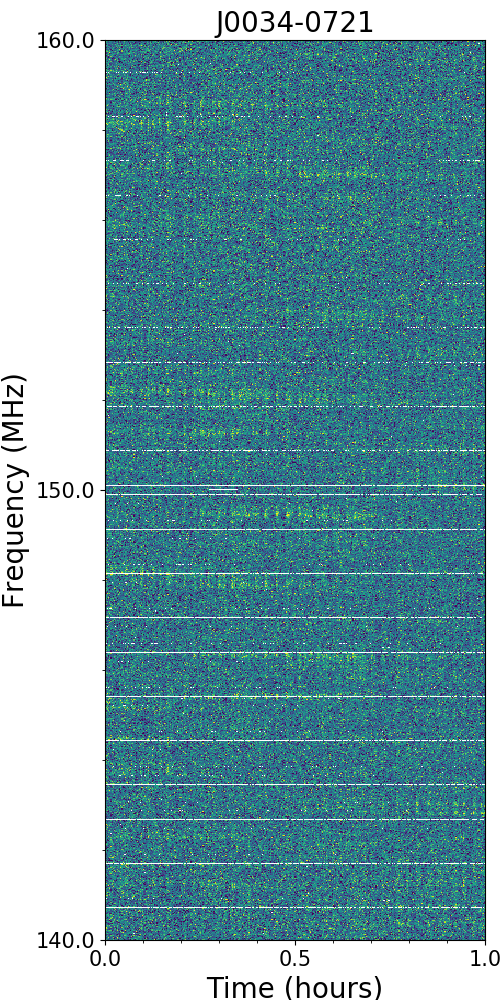}\hfill
     \includegraphics[width=0.195\linewidth]{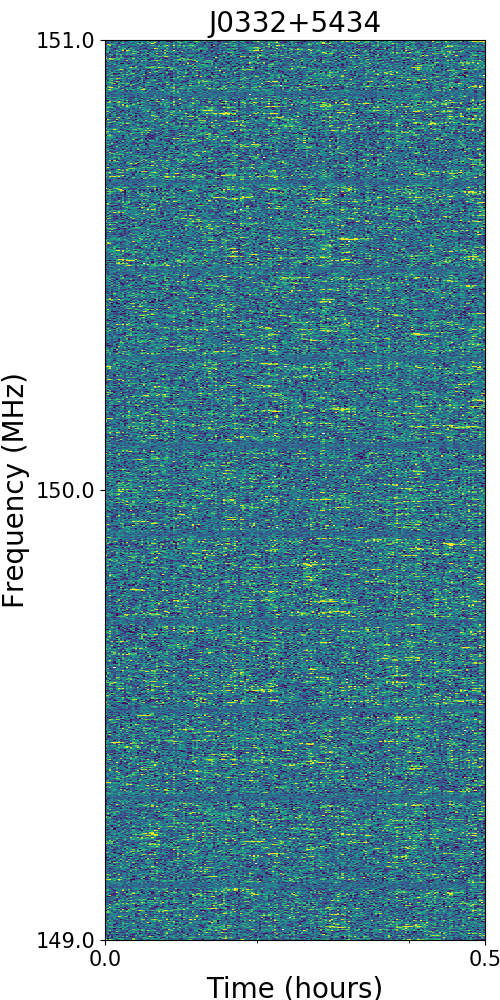}\hfill
     \includegraphics[width=0.195\linewidth]{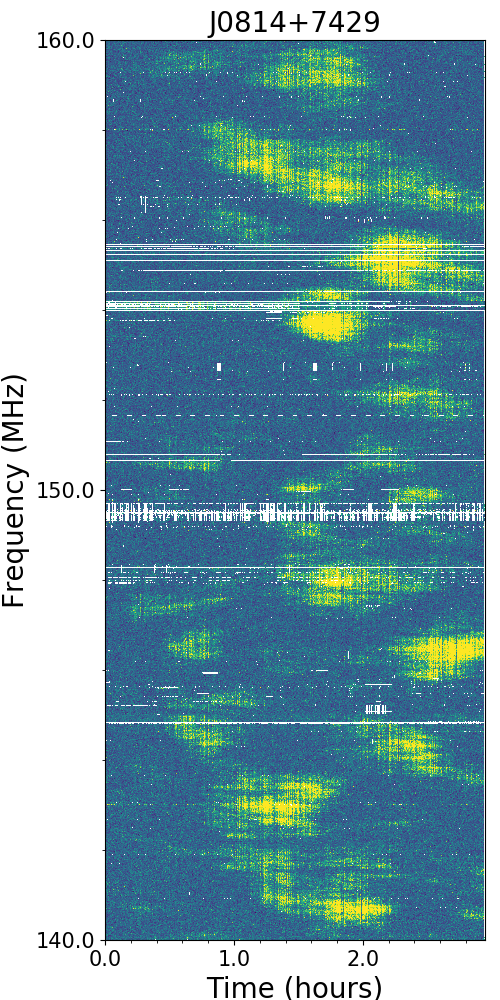}\hfill
     \includegraphics[width=0.195\linewidth]{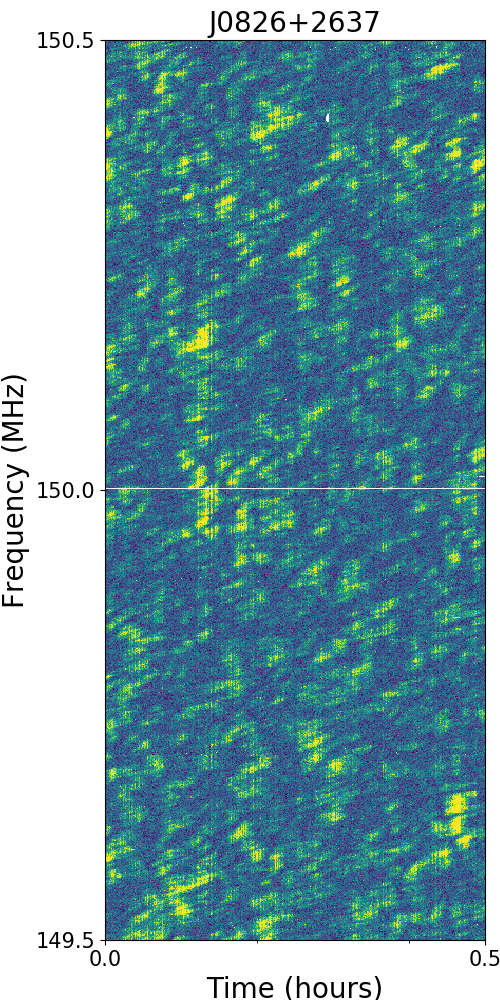}\hfill
     \includegraphics[width=0.195\linewidth]{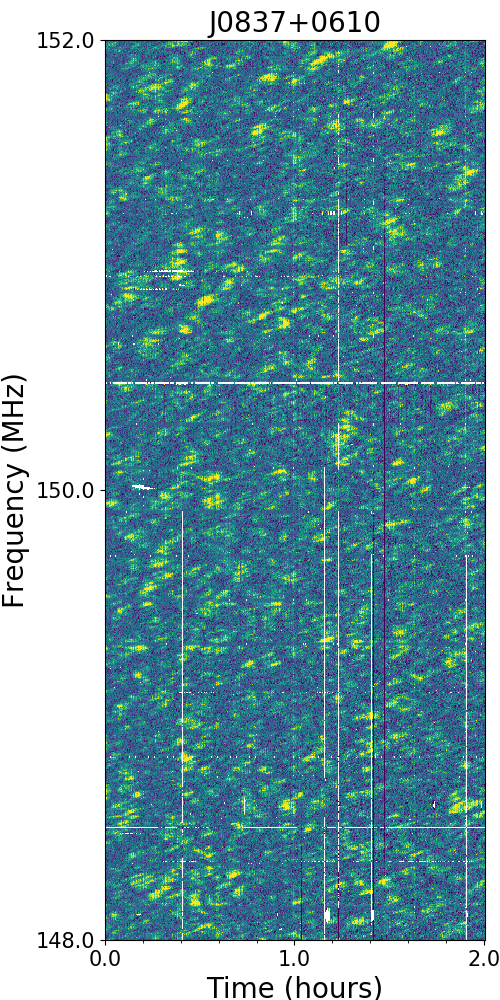}\hfill
     \includegraphics[width=0.195\linewidth]{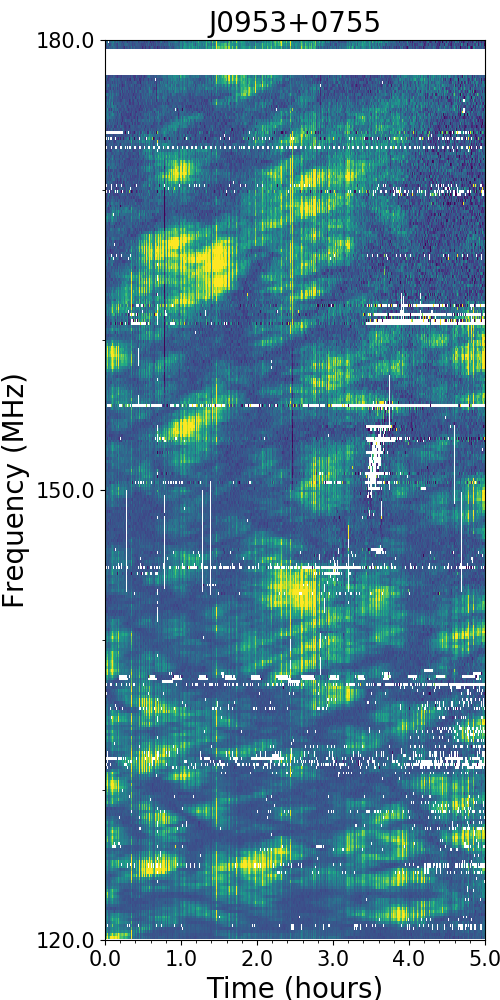}\hfill
     \includegraphics[width=0.195\linewidth]{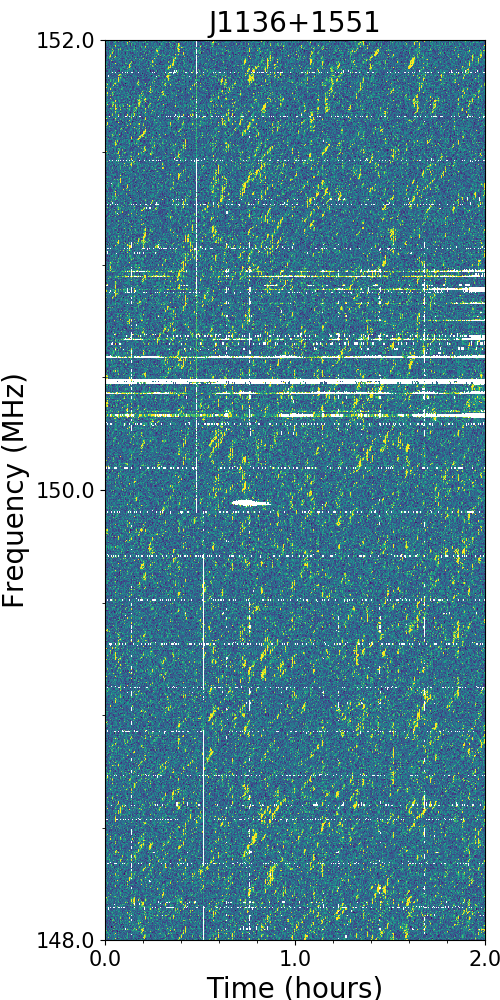}\hfill
     \includegraphics[width=0.195\linewidth]{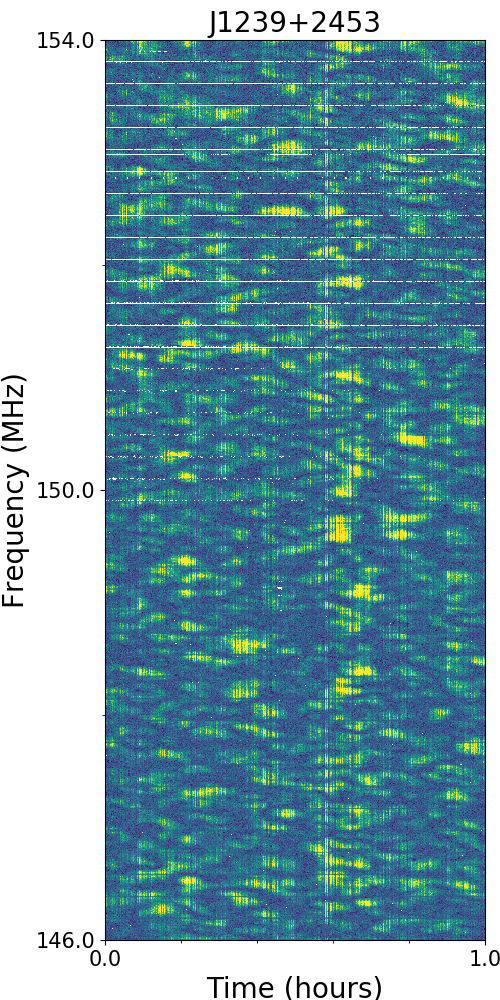}\hfill
     \includegraphics[width=0.195\linewidth]{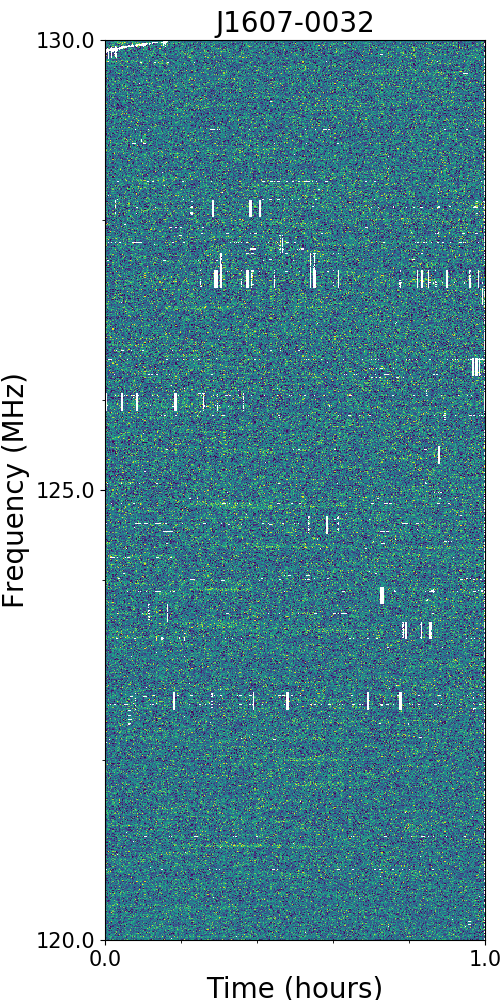}\hfill
     \includegraphics[width=0.195\linewidth]{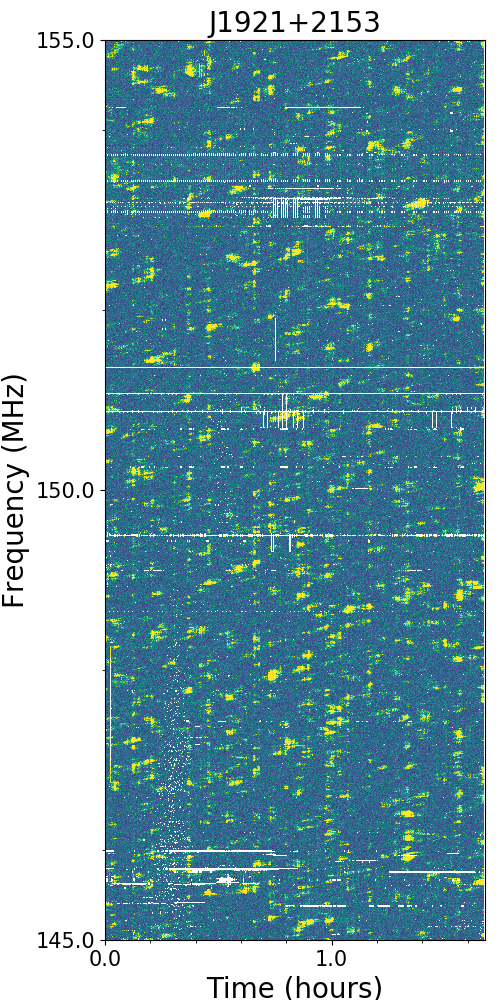}\hfill
     \includegraphics[width=0.195\linewidth]{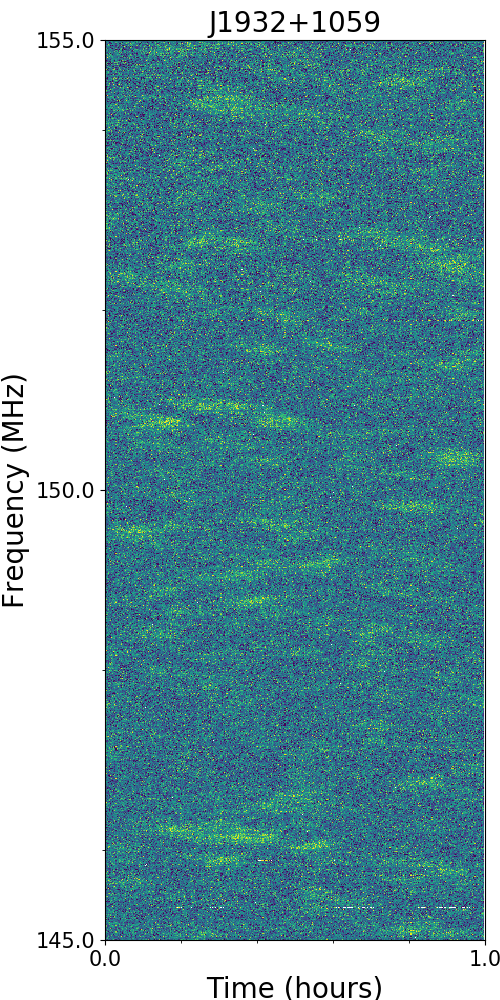}\hfill
     \includegraphics[width=0.195\linewidth]{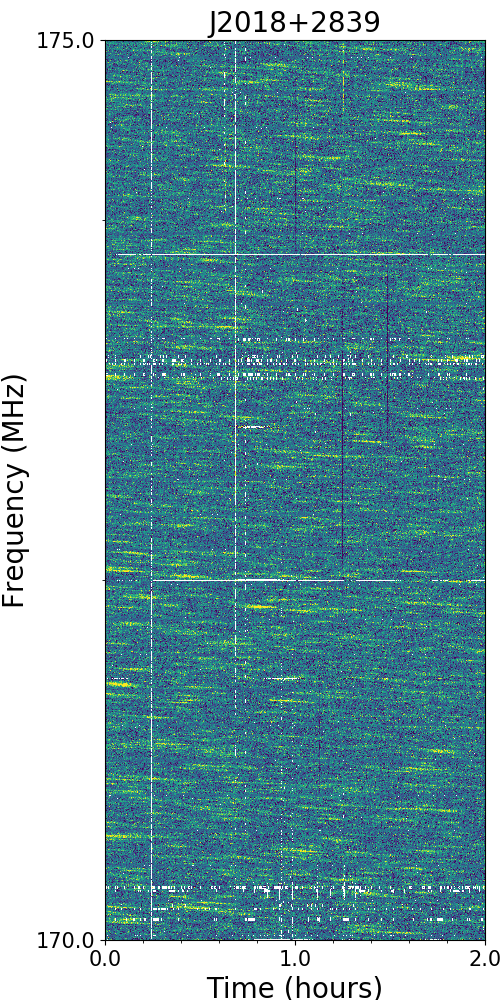}\hfill
     \includegraphics[width=0.195\linewidth]{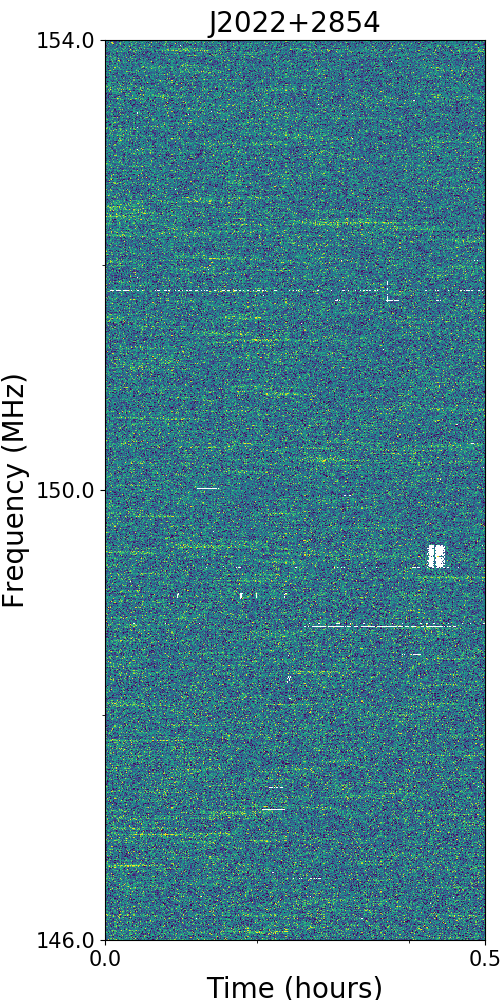}\hfill
     \includegraphics[width=0.195\linewidth]{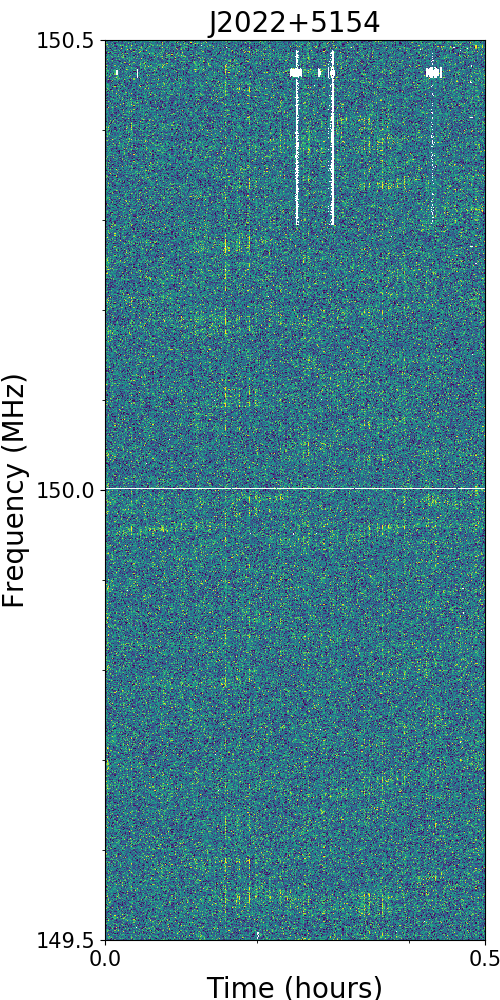}\hfill
     \includegraphics[width=0.195\linewidth]{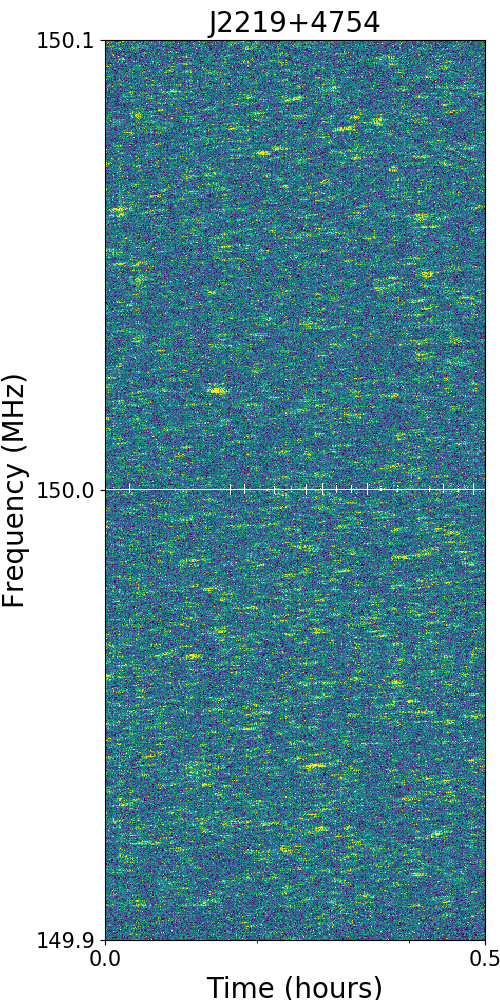}\hfill
      \caption{Dynamic spectra of 15 scintillating pulsars with
        LOFAR. The white patches were removed because of
        radio-frequency interference. The color scheme indicates the
        pulse S/N ranging from blue ($\overline{\rm{S/N}}$ - 2
        $\times$ $\sigma_{\rm{S/N}}$) to yellow ($\overline{\rm{S/N}}$
        + 3 $\times$ $\sigma_{\rm{S/N}}$), which is heavily modulated
        due to diffraction in the interstellar medium. The high-S/N
        "islands" are commonly referred to as scintles and provide
        information on the turbulent interstellar plasma, as described
        in the text.}
      \label{fig:scintillation_census}
    \end{figure*}

\begin{figure}[ht]
    \centering
    \includegraphics[width=0.49\linewidth]{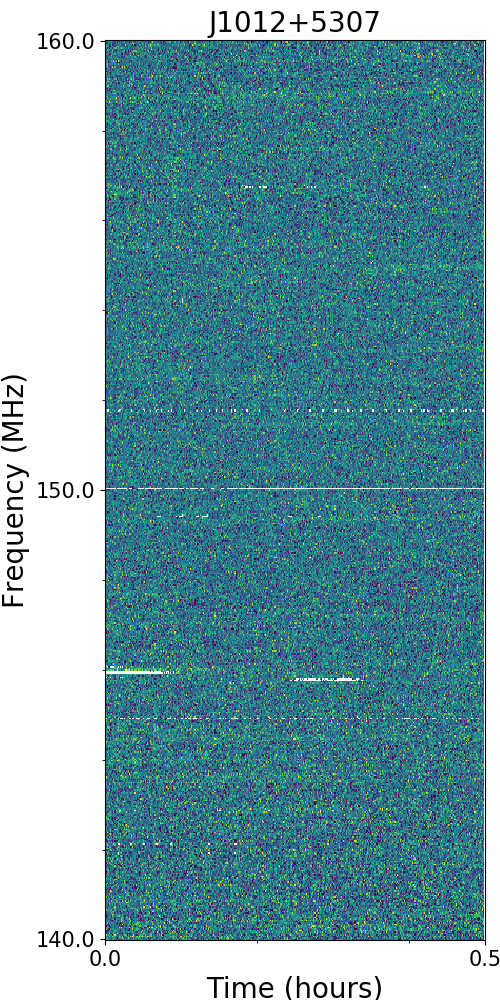}
    \includegraphics[width=0.49\linewidth]{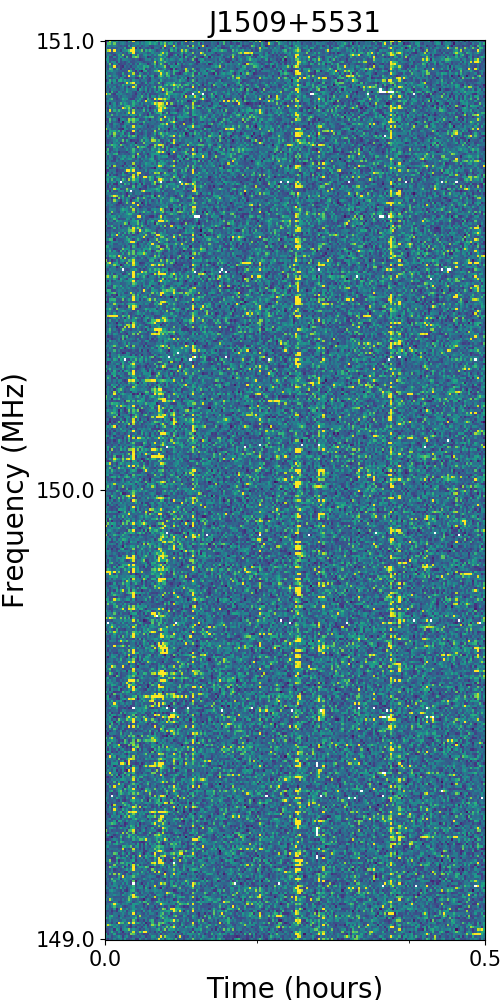} \\
    \caption{The dynamic spectra of PSRs~J1012+5307 and J1509+5531
      from the LOFAR Core and DE602, respectively, illustrating the
      challenges inherent to scintillation-detection at low
      frequencies. While scintillation is clearly visible in the
      PSR~J1509+5531 data, higher frequency resolution is required to
      allow accurate determination of the scintillation
      parameters. For PSR~J1012+5307 the resolution is sufficient, but
    the remaining S/N for individual scintles, is insufficient to
    allow clear analysis. While the dynamic spectrum does combine data
    from all scintles across the observation, this still returns a S/N
    that is too low for reliable measurements in this case. }
    \label{fig:dynamic_spectra_from_more_pulsars}
    \end{figure}

    From the 31 pulsars that were studied in this work, 15 showed
    clear scintillation at LOFAR frequencies (see
    Table~\ref{tab:properties}, Figure~\ref{fig:scintillation_census} and Figure~\ref{fig:scintillation_arc_census}), while 15 either did not show evidence
    for scintillation at all, or did show scintillation but with
    insufficient S/N to allow reliable quantification (see
    Table~\ref{tab:more_tested_pulsars}).
    We note that non-detections
    can be expected for two reasons, as illustrated in
    Figure~\ref{fig:dynamic_spectra_from_more_pulsars}. In most cases
    our frequency resolution does not suffice to resolve the scintles
    and hence does not allow detailed measurements of the
    scintillation bandwidth, as in the case of the observation of
    PSR~J1509+5531. While in principle this could be remedied by using
    higher frequency resolution, this is limited by available
    processing resources, by the amount of time resolution that is
    needed to resolve the pulse profile (particularly in the case of
    rapidly spinning MSPs) and by the S/N one can achieve in an
    extremely narrow bandwidth. In the case of PSR~J1012+5307, e.g.,
    the scintles appear to be resolved, but even with the full
    sensitivity of the LOFAR core they only barely stand out above the
    noise\footnote{The ACF combines the signal from all
      scintles within the observation, but in practice this is still
      insufficient for many sources.}.

    Representative dynamic spectra for the 15 pulsars with detected
    scintles are shown in Figure~\ref{fig:scintillation_census}. Clear
    criss-cross structures indicative of scintillation arcs, can be
    seen for example for PSRs~J0826+2637, J1136+1551 and J1932+1059, a few more
    arcs can be seen in the secondary spectra (see
    Figure~\ref{fig:scintillation_arc_census}), although these are in
    many cases hard to study at these frequencies, due to the fact
    that they tend to lie close to the delay axis and because they are
    far less sharp than at higher frequencies
    \citep{rszm21}. Nevertheless, several
    of these arcs are sufficiently clearly detected to allow detailed
    studies, even at these frequencies. 

\begin{figure*}[hbtp]
\subfloat{\includegraphics[width=0.31\linewidth]{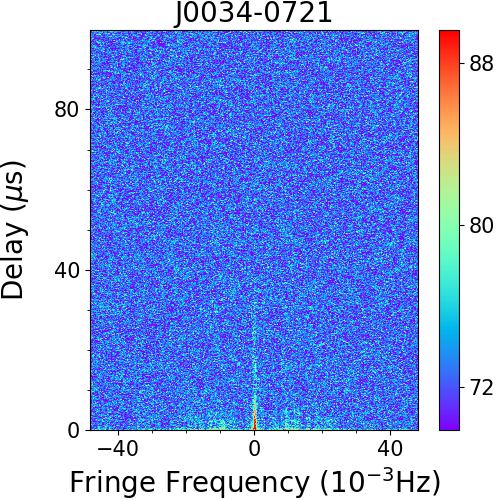}\hfill}
\qquad 
\subfloat{\includegraphics[width=0.31\linewidth]{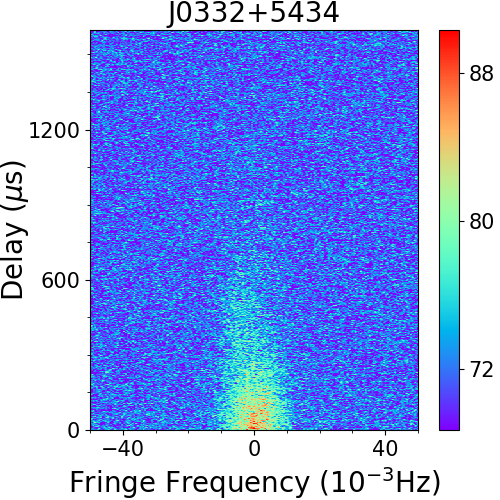}\hfill}
\qquad 
\subfloat{\includegraphics[width=0.31\linewidth]{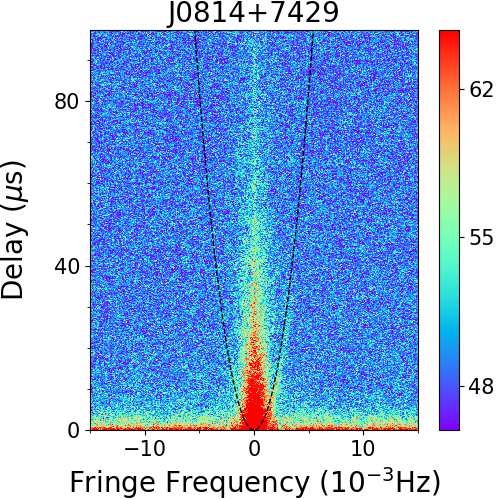}\hfill}
\qquad 
\subfloat{\includegraphics[width=0.31\linewidth]{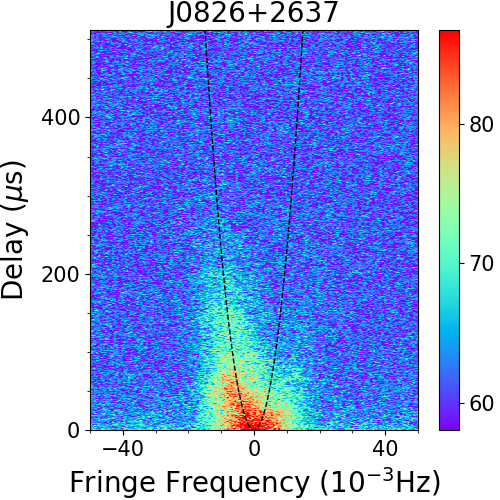}\hfill}
\qquad 
\subfloat{\includegraphics[width=0.31\linewidth]{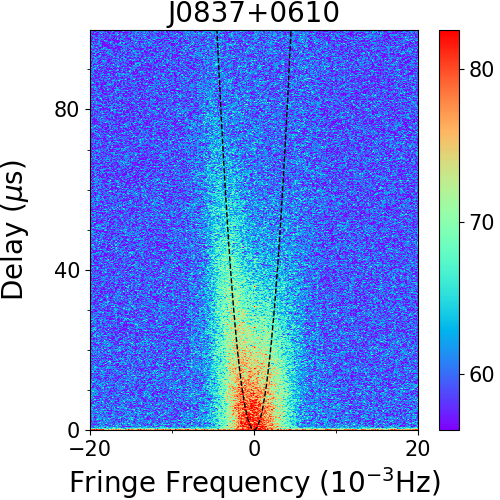}\hfill}
\qquad 
\subfloat{\includegraphics[width=0.31\linewidth]{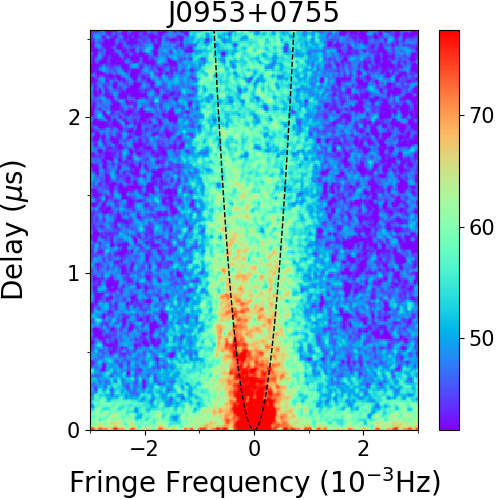}\hfill}
\qquad 
\subfloat{\includegraphics[width=0.31\linewidth]{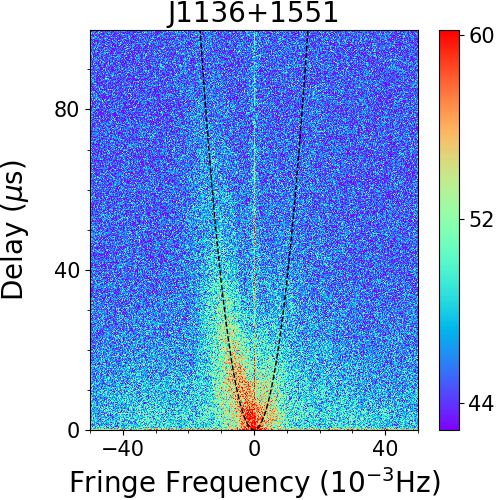}\hfill}
\qquad 
\subfloat{\includegraphics[width=0.31\linewidth]{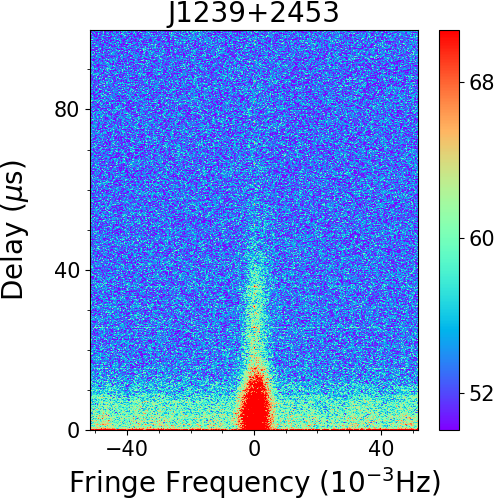}\hfill}
\qquad 
\subfloat{\includegraphics[width=0.31\linewidth]{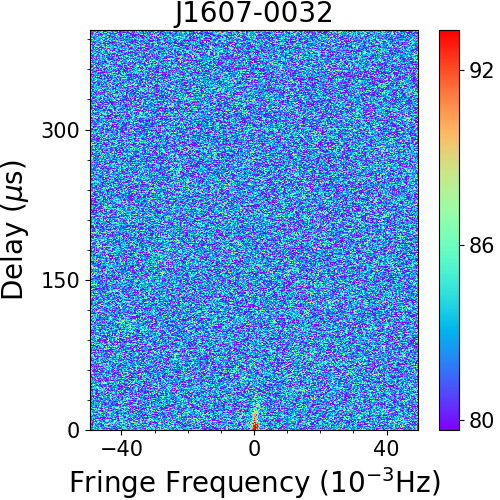}\hfill}
\qquad 
\subfloat{\includegraphics[width=0.31\linewidth]{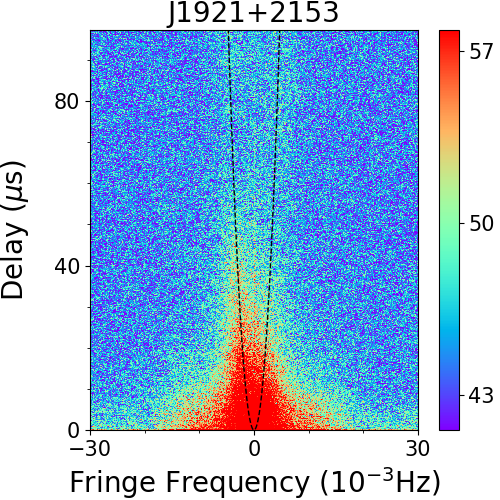}\hfill}
\qquad 
\subfloat{\includegraphics[width=0.31\linewidth]{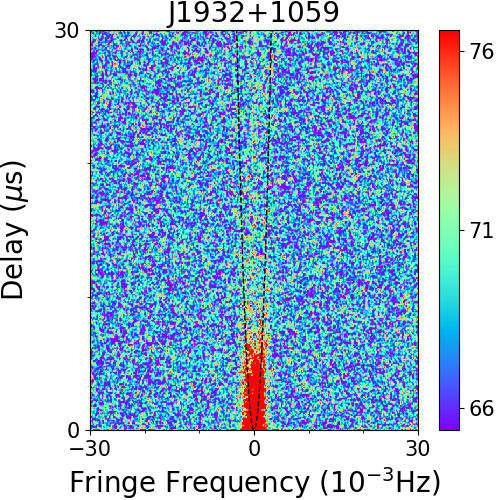}\hfill}
\qquad 
\subfloat{\includegraphics[width=0.31\linewidth]{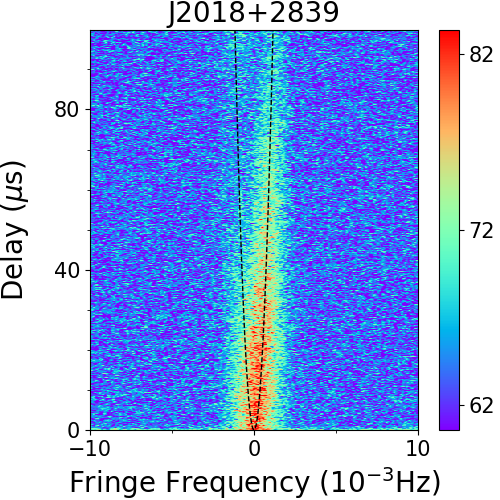}\hfill}
\caption{Secondary spectra of 15 scintillating pulsars detected
        with LOFAR in the frequency range 145-155 MHz (except for
        PSR~J0814+7429: 125$\pm$5 MHz; PSR~J0953+0755: 150$\pm$20 MHz and
        PSR~J2018+2839: 175$\pm$5 MHz). The black parabolae are the
        resulting model fits with parameters given in
        Table~\ref{tab:arc}.}
\label{fig:scintillation_arc_census}
\end{figure*}

\begin{figure*}[hbtp]
\ContinuedFloat 
\qquad 
\subfloat{\includegraphics[width=0.31\linewidth]{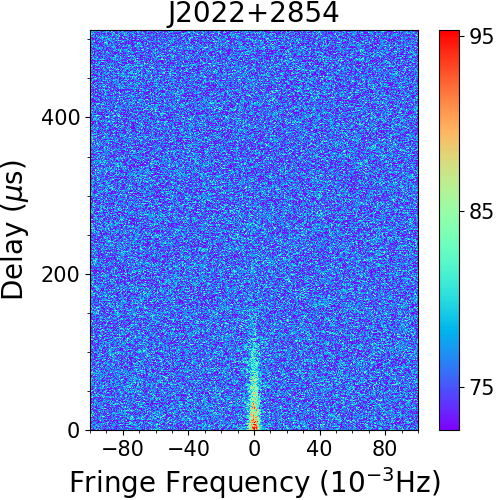}\hfill}
\qquad 
\subfloat{\includegraphics[width=0.31\linewidth]{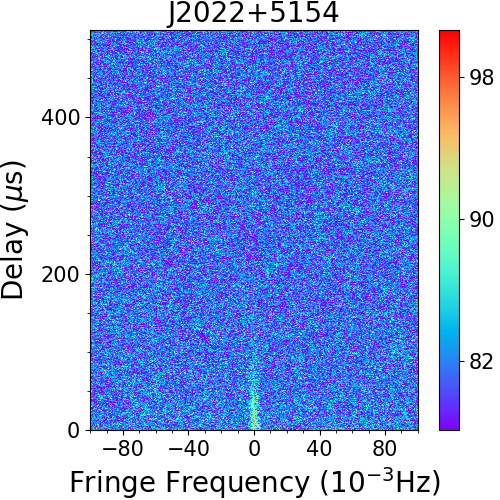}\hfill}
\qquad 
\subfloat{\includegraphics[width=0.31\linewidth]{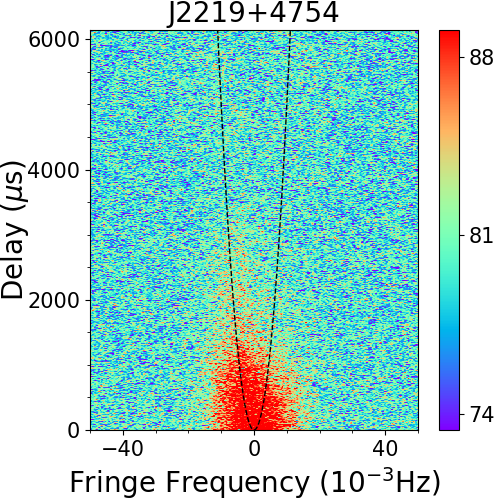}\hfill}
\caption{Continued}
\end{figure*}

\begin{figure}[ht]
      \centering
      \includegraphics[width=0.49\linewidth]{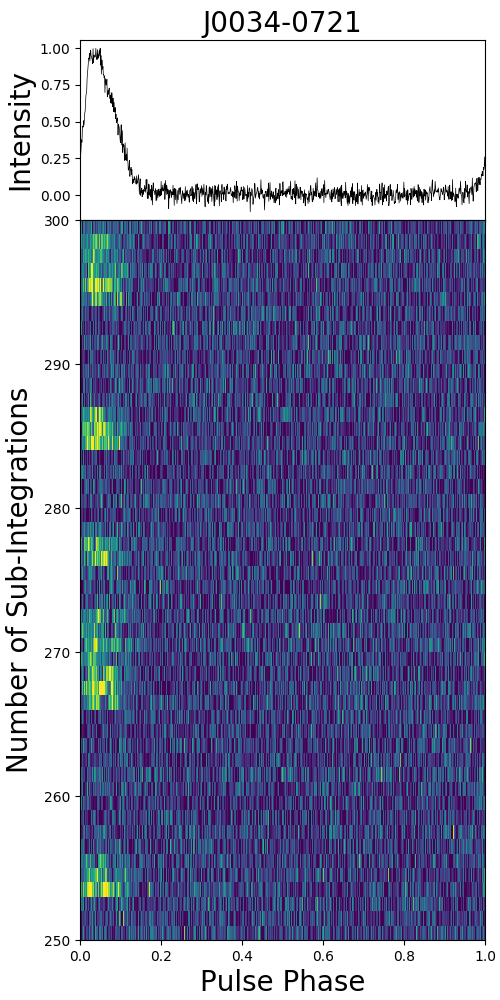}
      \includegraphics[width=0.49\linewidth]{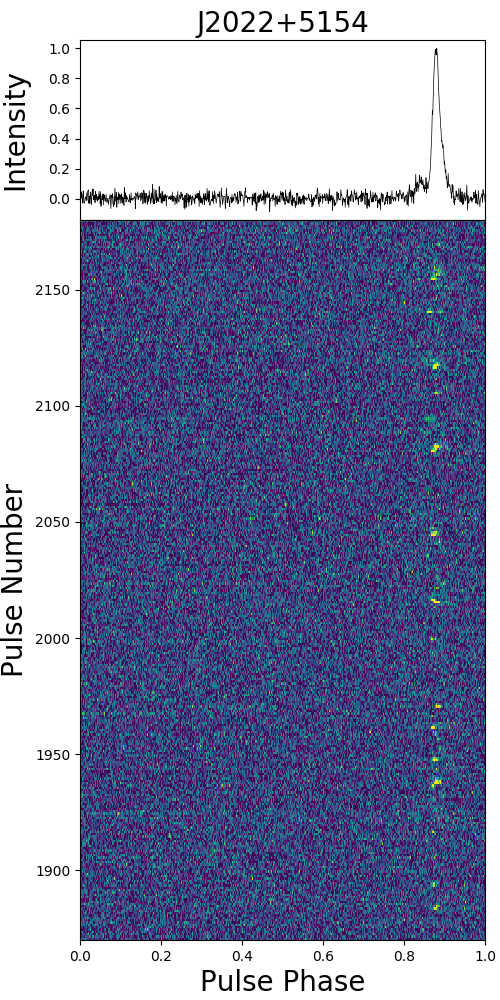}
      \caption{Nulling behaviour of pulsars in our
        sample. PSRs~J0034$-$0721 and J2022+5154 show nulling
        behaviour as shown above. The top row shows the integrated
        profile of a 1-hour observation with FR606 across the
        frequency range 125-150 MHz for PSR J0034$-$0721 (left) and of
        a half-hour observation with the LOFAR core across the
        frequency range 149-153 MHz. The intensity-phase plots of
        these data are shown in the bottom row as a function of the integration
        number for the PSR~J0034$-$0721 data (which has a 10-second
        sub-integration length) and as a function of the pulse number
        for the single-pulse data on PSR~J2022+5154. The intermittent
        character of their emission makes determination of the
        scintillation timescale complicated, but the scintillation
        bandwidth can still be determined using standard methods. 
      } 
      \label{fig:nulling}
    \end{figure}

    Two of the pulsars (PSRs~J0034$-$0721 and J2022+5154) in our sample display pulse-nulling behaviour
    (see Figure~\ref{fig:nulling}), which can be seen as vertical
    structure in Figure~\ref{fig:scintillation_census}, which could be
    removed by applying a Wiener filter to the dynamic spectrum, as
    described by \citet{lmv+21}, but this is beyond the scope of the
    present paper. While this nulling prevents accurate determination
    of $\tau_{\rm d}$, we can still determine $\Delta \nu_{\rm d}$ 
    without a problem. 
    Nevertheless, the uncertainty determination for $\Delta \nu_{\rm d}$ is problematic since the statistical error $\sigma_{\rm{est}}$ given in Eq.~\ref{eq:error} contains an unreliable measurement of $\tau_{\rm{d}}$ from these two nulling pulsars.     
    We note that the $\rm{BW_{\rm{dyn}}/\Delta\nu_{\rm{d}}}$ term plays a major role in deriving the statistical error $\sigma_{\rm{est}}$.

    From our measured scintillation parameters, some further measures
    can be derived to describe the IISM, as discussed in detail by
    \citet{r90}. Specifically, the ratio of the Fresnel scale and the
    DISS scale\footnote{The DISS scale or the spatial scale of
      diffractive scattering, $s_{\rm d}$, is defined as the
      transverse separation within which incident waves have an root mean square
      phase difference of one radian or less.} $s_{\rm d}$, quantifies
    the scattering strength $u$ \citep{r90,brg99}:
    \begin{equation}
      u \approx \sqrt{\frac{2\nu}{\Delta \nu_{d}}}.
	  \label{eq:scintillation strength}
    \end{equation}
    The observed values of $u$ from 15 pulsars are given in column 10
    of Tab.\ref{tab:properties}, showing that all our observations are
    clearly in the strong scattering regime ($u>1$). Within this
    regime, we can estimate the time-scale of refractive scintillation
    (RISS), using \cite{r90,wmj+05}:
    \begin{equation}
      t_{\rm{r}} \approx \frac{2\nu}{\Delta \nu_{d}} \tau_{\rm{d}}.
	  \label{eq:ref_sci_time_scale}
    \end{equation}
    The derived values of $t_{\rm{r}}$ are shown in column~11 of
    Table~\ref{tab:properties} for LOFAR frequencies; under
    the assumption of a Kolmogorov turbulence spectrum, these scale as
    $t_{\rm r} \propto \nu^{-2.2}$ with observing frequency. Given
    that our observation lengths are orders of magnitudes smaller than
    $t_{\rm r}$ for all pulsars in our sample, we can be confident our
    DISS measurements are uncorrupted by RISS effects. Note, however,
    that $\tau_{\rm d}$ is strongly affected by the relative motion of
    the Earth and the pulsar and can hence affect $t_{\rm r}$
    significantly throughout the year, especially for pulsars with
    relatively small transverse velocities. Since our analysis is
    based on single-epoch observations, our results remain unaffected
    by this.
    
    \subsection{Turbulence characteristics of the IISM}
    Between the inner and outer scales of turbulence \citep[the so-called
    inertial subrange scales,][]{zs98} electron-density fluctuations follow the well-known Kolmogorov
    spectrum \citep{ars95}. 
    The inner scale is constrained
    to $\sim$100 km \citep{sg90, rjt+09} and the outer scale is
    of order $\sim$100 pc \citep{ars95, xz17} or 1-20 pc
    \citep{rjt+09}. It is also increasingly clear that the underlying
    scattering structures are often anisotropic \citep{bmg+10,wdb+09,sro19}.
    
    Our data allow a meaningful test of the turbulent spectrum of the
    IISM, since at these low frequencies, even a narrow frequency
    range can obtain numerous scintles (see
    Figure~\ref{fig:scintillation_census}), enabling a self-consistent
    instantaneous test of the frequency-scaling laws in the turbulent
    medium. Specifically, $\Delta \nu_{\rm d}$ is predicted to scale
    with the observing frequency as 
    $\Delta\nu_{\rm{d}} \propto \nu^{-\alpha}$, where the power-law
    index $\alpha = 4.4$ for a Kolmogorov turbulence spectum and
    $\alpha$ = 4.0 for Gaussian turbulence. An alternative method to
    determine $\alpha$ is by measuring the scatter-broadening, as done
    by \citet{bts+19, kmj+19, gkk+17}, e.g.

\begin{figure*}
      \centering
      \includegraphics[width=0.99\linewidth]{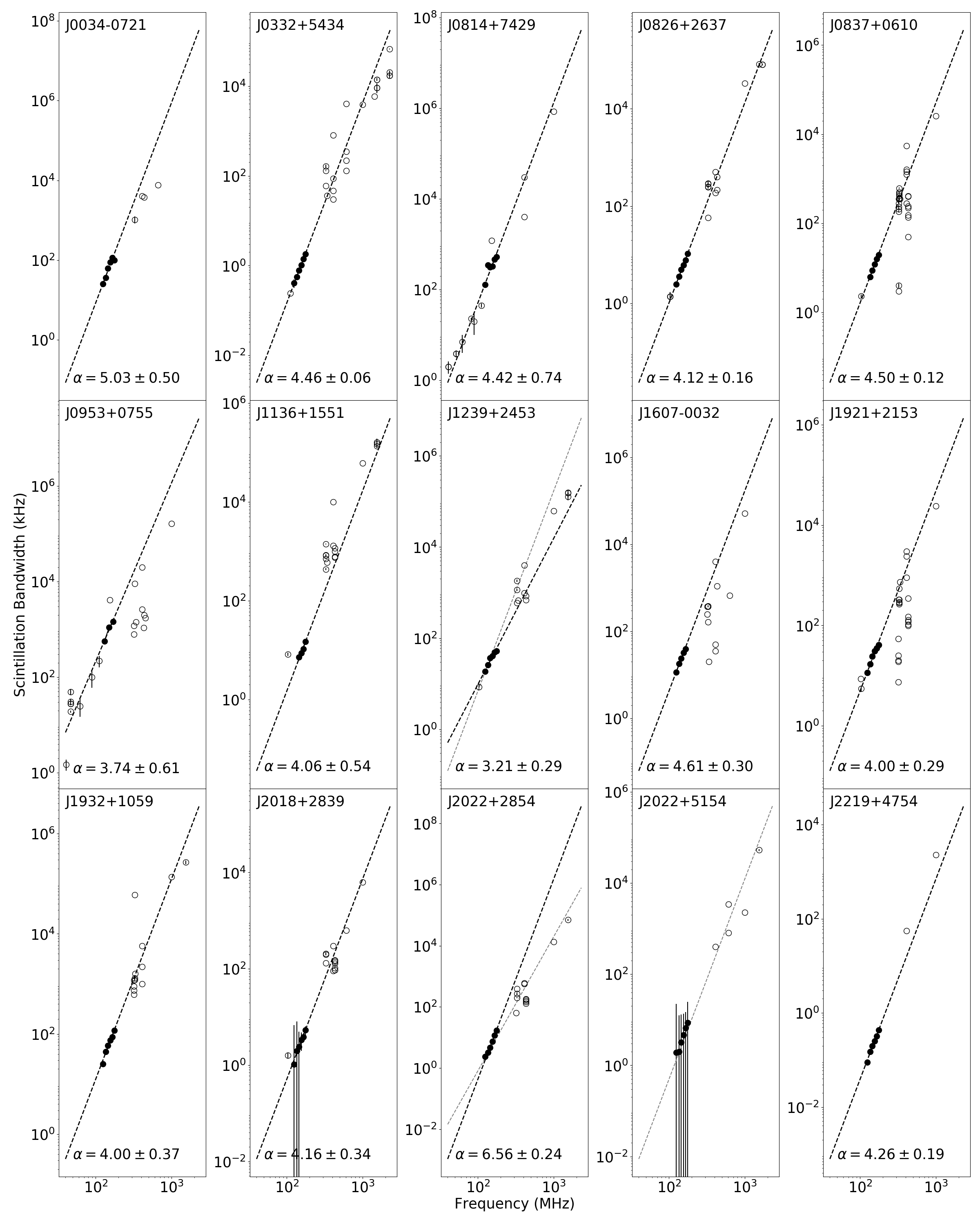}
      \caption{The scintillation bandwidth versus the observing
        frequency in log-log scale. Filled symbols designate our LOFAR measurements presented in this paper, open symbols are previously published values collated from literature and given
         (with references) in the Table in Appendix~\ref{appendix:a}. 
         The black dashed lines with a slope
        of $\alpha$ show the best fit to the $\Delta \nu_{\rm{d}}$
        values across our LOFAR observing band. The gray dashed line
        corresponds to the Kolmogorov spectrum with $\alpha$ = 4.4
        (which is only plotted in case the best-fit line is
        inconsistent with a Kolmogorov spectrum).}
      \label{fig:scintillation_frequency_rescaling}
    \end{figure*}

    \paragraph{Determination of $\alpha$.}
    In this work, we report independent measurements of $\alpha$
    from our LOFAR data set. We obtain $\Delta\nu_{\rm{d}}$ from
    dynamic spectra of 10-MHz-wide frequency bands for each pulsar,
    except for PSR~J0953+0755, in which case we adopt a 20-MHz-wide
    band in order to increase the number of scintles and thus reduce the statistical
    uncertainty. Then, based on the measurements of $\Delta\nu_{\rm{d}}\left(\nu\right)$
    (obtained from a single observation), we are able to get the
    parameter $\alpha$ without any influence from RISS. The
    measurements of $\Delta \nu_{\rm{d}}$ for the 15
    nearby pulsars in our sample are given along with any previously
    published measurements of $\Delta \nu_{\rm{d}}$ in
    Appendix~\ref{appendix:a} and are shown in
    Figure~\ref{fig:scintillation_frequency_rescaling}. The derived
    values for $\alpha$ are listed in Table~\ref{tab:properties}. The large
    uncertainty of $\Delta\nu_{\rm{d}}$ for 
    PSR~J2022+5154
    is due to an insufficient S/N that makes the 2D ACF ill-defined,
    preventing a measurement of $\alpha$. 
    
    \paragraph{Comparison with Theory.}
    The measured $\alpha$ values are mostly consistent with the predictions for
    both a Kolmogorov and a Gaussian spectrum, except for
    PSRs~J1239+2453 ($\alpha = 3.21 \pm 0.29$) and J2022+2854 ($\alpha
    = 6.56 \pm 0.24$), which are too shallow and too steep
    respectively. For convenience, the gray dashed lines that
    represent the prediction for $\alpha=4.4$ are also presented for
    these two pulsars in
    Figure~\ref{fig:scintillation_frequency_rescaling}. For
    PSR~J2022+2854, the Kolmogorov spectrum does match previously
    published values of $\nu_{\rm d}$ across a wider frequency range,
    even if within our LOFAR band a much steeper gradient is observed.
    In the case of PSR~J1239+2453, previously published values for
    $\nu_{\rm d}$ are scattered and do not seem to prefer either the Kolmogorov gradient or the shallower spectrum derived from our LOFAR data.  

    A flatter $\alpha$ may be due to associations with H{\textsc II}
    regions, spiral arms or supernova remnants \citep{gn85}, finite and
    anisotropic scattering screens \citep{cl01,gkk+17} or 
    the fact that the diffraction scale $s_{\rm{d}}$ becomes
    smaller than the inner scale at lower frequencies
    \citep[e.g.][]{bts+19}. It has also been suggested \citep{xz17}
    that a composite electron-density spectrum could cause shallower
    values of $\alpha$ for high-DM pulsars.

   For PSR~J1239+2453,
    $s_{\rm{d}}\sim1/(k\theta_{\rm{d}})$ $\sim$ 31000\,km which is
    larger than the inner scale, where $\theta_{\rm{d}}$ is the width
    of the angular scattering given by $\sqrt{c/\pi
      D_{\rm{p}}\Delta\nu_{\rm{d}}}$ $\sim 2.1$ mas in which
    $D_{\rm{p}}$ = 0.85$\pm$0.06 kpc \citep{bbg+02} and $k$ is the
    wave number, following \citet{r90}. Consequently, the inner scale
    cannot be the cause for the shallow spectrum we observe; also the
    suggestion of \citet{xz17} is not relevant to this pulsar, as it
    lies above the Galactic plane ($G_{\rm b}$ = $86.5^{\circ}$) and
    has a small DM of 9.25\,pc/cm$^3$. Significant anisotropy in the
    scattering medium may well provide an explanation for the shallow
    spectrum, as the power-law indices of pulsars with scintillation
    arcs have been observed to be shallower \citep{sro19} and these
    arcs are caused by anisotropies. For PSR~J1239+2453 this could be
    a possible explanation, since arcs have recently been detected at
    higher frequencies \citep{fab+18}, even if they are
    undetectable in our LOFAR data. 

    For PSR~J2022+2854, the steep $\alpha$ could be because of scattering from large-scale inhomogeneities involved, which may suggest that the effect of
    refractive scattering is large and is independent of observing
    frequency and the pulsar distance \citep{gn85}.  
    This requires further investigation of
    RISS and is deferred to a future paper. 

    \paragraph{Comparison with Literature.}
    The $\alpha$ for PSRs~J0332+5434, J2018+2839 and
    J2219+4754 agree with the values reported by \citet{kjm17, bts+19,
      kmj+19}.  However, our $\alpha$ measurement of PSR~J0826+2637
    ($4.1\pm0.2$) strongly differs with those in literature:
    1.55$\pm$0.09 \citep{bts+19} and 2.4$\pm$0.1 \citep{kmj+19}.  This
    could result from the frequency-dependent evolution of $\alpha$
    \citep{sg90} and will be investigated in detail in the second
    paper in this series. 
    For the remaining pulsars in our sample, no previous
    estimates of $\alpha$ were published to date, which is partly
    expected since the pulse scatter broadening becomes unresolvably
    small, particularly at higher frequencies. In this regard the
    recent NenuFAR upgrade to the LOFAR station in Nan\c{c}ay
    observatory has great potential to complement our work
    \citep{bgt+21}.
      
    With $\alpha$, we are able to predict scintillation bandwidth at
    other frequencies (see
    Figure~\ref{fig:scintillation_frequency_rescaling} and
    Appendix~\ref{appendix:a}).  The $\Delta \nu_{\rm{d}}$ values from
    literature are typically consistent with our predictions, although
    some discrepancies exist. Specifically, there are significant
    differences for PSRs~J0837+0610, J1607$-$0032 and J1921+2153. The
    previously published measurements of scintillation bandwidth for
    these three sources span a wide range from $\sim$100 MHz to
    1\,GHz. While even for these three pulsars most previously
    published values are consistent with our predictions,
    in particular across the 300-500 MHz range there are some significant discrepancies. Below, we discuss the potential causes of these
    discrepancies.

\begin{enumerate}
  \item Modulation by RISS. \cite{bgr99} presented long-term
  scintillation observations of 18 pulsars, whereby the variation of $\Delta
  \nu_{\rm{d}}$ was clearly detected for all pulsars but could only be
  explained in terms of RISS for two pulsars in their sample.
  However, traditional scintillation theory is based on an infinite and
  isotropic scattering screen, whereas presently it is well established that
  highly anisotropic scattering screens are common
  \citep{smc+01,bmg+10}. This implies that more extreme $\Delta \nu_{\rm d}$
  variations than predicted from standard theory may still be caused by RISS, although the possible amplitude of such variations is ill-defined. 
  \item Limited bandwidth and observing length. The limited window of
    measured dynamic spectra can cause large differences due to the
    small number of scintles, particularly at high observing
    frequencies for low DM pulsars. A clear discussion of this was
    recently published by \cite{bgp+22}.
  \item Short-lived discrete ionised clouds. 
  During the epochs arclets existed, the scintillation bandwidth
  $\Delta \nu_{\rm{d}}$ always exhibits considerably smaller values,
  e.g., 4.0 $\pm$ 0.5\,kHz at 324\,MHz and $\sim$3\,kHz at 327\,MHz
  for PSR~J0837+0610, as reported by \cite{ssa+20} and \cite{bmg+10}. (Note that
    we predict a value of $\sim$325\,kHz for $\Delta \nu_{\rm{d}}$ at
    those frequencies). 
    The observable arclets imply that in addition to the central
    screen, discrete clumps contribute to scattering as well
    \citep{crs+06}, which equivalently means a high scattering angle
    is involved, corresponding to a smaller scintillation bandwidth $\Delta \nu_{\rm{d}}$.
    We anticipate the detection of variation in scintillation
    bandwidth $\Delta \nu_{\rm{d}}$ as such discrete clouds pass
    the line of sight. 
  \item Overall variability in the IISM. In addition to extreme events
    like those proposed by \citet{ssa+20}, more gradual changes in the
    IISM along the line of sight could cause evolution of
    scintillation parameters on longer time scales, leading to
    differences between values that are published many years
    apart. \citet{bts+19} and \citet{bgr99} studied the temporal
    evolution of scintillation parameters on time scales of years and
    showed that while such secular variations do occur along some
    lines of sight, in most cases they are either absent or very
    shallow. 
  \item Non-astronomical causes. 
    There are other possible
    reasons such as RFI, instrumental failures, general instrumental
    limitations or potential errors in procedures applied in obtaining
    the scintillation parameters. 
\end{enumerate}

    \subsection{Scintillation arcs}
    
    \begin{table*}[ht]
    \begin{threeparttable}
      \centering
      \caption{Scintillation-arc properties for nine pulsars with
        clearly detected scintillation arcs in our LOFAR data.}
      \begin{tabular}{cccllccl}
        \toprule \toprule
        PSR        & MJD   & Freq.    & $\eta$   & \multicolumn{1}{c}{$L_{s, \parallel}$}&$D_{\rm s_{low}}$&$D_{\rm{s}}^{*}$ from                & $D_{\rm{p}}$        \\ 
        Name       &       & (MHz)    & ($s^{3}$)& \multicolumn{1}{c}{(au)} & (pc)               & literature (pc)                         & (kpc)               \\          
        \midrule                                                                  
        J0814+7429 & 57872 & 120--130 & 3.3(1)   & \phantom{$\ge$}0.05      & 350(13)            & uniform or 275--355 $^{a}$ & 0.433(8)$^{\rm{b}}$ \\
        J0826+2637 & 58820 & 145--155 & 2.3(7)   & \phantom{$\ge$}0.19      & 180(50)            & (140, 190, 375, 470)$^{c}$ & 0.5$^{\rm{d}}$    \\
        &&&&&& 240$\pm$90$^{e}$ & \\
        J0837+0610 & 58867 & 145--155 & 4.90(2)  & \phantom{$\ge$}0.12      & 330(30)            & 460$\pm$80$^{f}$,  420$^{g}$, 410$\pm$40$^{h}$   & 0.62(6)$^{i}$       \\
        &&&&&& 420$\pm$90$^{e}$ & \\
        J0953+0755 & 57391 & 130--170 & 4.8(7)   & $\ge$0.01                & 230(35)            & 4.4–16.4 and 26–170 $^{\rm{j}}$     & 0.262(5)$^{b}$            \\
        J1136+1551 & 57122 & 145--155 & 0.37(3)  & \phantom{$\ge$}0.09      & 120(12)            & (21, 56, 136, 189)$^{c}$, $\geq$136$^{k}$                           & 0.35(2)$^{\rm{b}}$  \\
        J1921+2153 & 58356 & 145--155 & 4.4(1)   & $\ge$0.14                & 500(15)            & uniform or 360$^{l}$                 & 0.81$^{\rm{m}}$     \\
        J1932+1059 & 59213 & 145--155 & 3.0(1)   & \phantom{$\ge$}0.04      & 180(10)            &  (73, 202, 301)$^{c}$,    & 0.33(1)$^{\rm{b}}$  \\
         &  &  &    &      &             & 200$\pm$20$^{n}$, 240$\pm$30$^{e}$, 190$\pm$50$^{e}$   &   \\
        J2018+2839 & 58844 & 170--180 & 74(5)    & $\ge$0.15                & 500(60) or 390(45) & $\le$100 or 550$\pm$30 $^{e}$                            & 0.95(9)$^{\rm{b}}$  \\
        J2219+4754 & 58863 & 145--155 & 50(9)    & \phantom{$\ge$}--        &    --              & --                                        & 2.39$^{\rm{m}}$     \\
        \bottomrule 
      \end{tabular}  
      \label{tab:arc}
      \begin{tablenotes}
      \small
      \item {\bf Notes:}  Given
        are
        the pulsar name,
        the date of the observation from which the arc parameters were derived,
        the range of frequencies,
        the arc curvature $\eta$,
        the spatial scale of the scattering screen $L_{\rm s, \parallel}$, 
        the lower limit on the distance between the scattering screen and Earth $D_{\rm s_{low}}$,
        any previously pulished estimates for the screen distance $D_{\rm s}$,
        and
        the assumed pulsar distance $D_{\rm p}$.
        Numbers in brackets give the nominal 1-$\sigma$ uncertainty in
        the last digit quoted and references are given as
        superscripted letters and are listed below the table. (Screen
        distances from literature were corrected for potential
        mismatches in pulsar distances used and are hence directly
        comparable with our results.)
      \item References: ($^{a}$) \citealt{rcm00}, ($^{b}$) \citealt{bbg+02}, ($^{c}$) \citealt{ps06a}, ($^{d}$) \citealt{dgb+19}, ($^{e}$) \citealt{fab+18}, ($^{f}$) \citealt{hsa+05}, ($^{g}$) \citealt{bmg+10}, ($^{h}$) \citealt{ssa+20}, ($^{i}$) \citealt{lpm+16}, ($^{j}$) \citealt{ssp+14}, ($^{k}$) \citealt{sro19}, ($^{l}$)\citealt{ssg+17}, ($^{m}$) \citealt{ymw17}, ($^{n}$) \citealt{yzw+20}.
     \end{tablenotes}
    \end{threeparttable}
    \end{table*}

    Scintillation arcs are well-described with the thin-screen approximation
    \citep{wms+04, crs+06}.
    The interfering unscattered rays at the piercing point of the
    direct line of sight towards the pulsar and scattered rays an angle $\theta$
    away from the piercing point 
    display a differential
    geometric time delay $\tau$ and differential Doppler shift
    $f_{\rm{D}}$, which causes the interference to be well-described as a
    parabola with curvature $\eta$ in the secondary spectrum, since $\tau$ and
    $f_{\rm D}$ relate as follows: 
    \begin{equation}
      \tau = \eta f_{\rm{D}}^{2},
	  \label{eq:arc_define}
    \end{equation}
    where the arc curvature $\eta$ can be shown to be defined as follows:
    \begin{align}
	  \label{eq:arc_calculate}
      \eta & = \frac{D_{\rm{p}}s(1-s)}{2\nu^{2}} \frac{c}{(\vec{V_{\rm{eff}}} \cos\psi)^{2}} \\
      & = 4.629 \times 10^{3} \frac{D_{\rm{p,kpc}}s(1-s)}{\nu_{\rm{GHz}}^{2}(\vec{V_{\rm{eff, km}}} \cos\psi)^{2}} \quad  (s^{3}).  \nonumber
    \end{align}
    Here $\vec{V_{\rm eff}}$ is the effective transverse line-of-sight
    velocity vector at the scattering screen, $c$ is the speed of light, $\nu$ is the observing
    frequency, $s$ is the fractional distance of the scattering
    screen, where $s=0$ is a screen at pulsar's location and $s=1$ is
    a screen at the Earth's position, $D_{\rm p}$ is the distance
    between the pulsar and Earth and $\psi$ is the angle between the
    (one-dimensional) scattering structure and $\vec{V_{\rm eff}}$ \citep{crs+06, yzw+20}. 

    For solitary pulsars, the effective transverse line-of-sight
    velocity $\vec{V_{\rm{eff}}}$ contain three components: the
    pulsar's transverse velocity $\vec{V_{\rm{p}}}$, the earth motion
    $\vec{V_{\rm{E}}}$ and the movement of scattering material
    $\vec{V_{\rm{ISM}}}$, which are combined as follows \citep{rch+19}:
    \begin{equation}
      \vec{V_{\rm{eff}}} = (1-s) \vec{V_{\rm{p}}} + s \vec{V_{\rm{E}}} - \vec{V_{\rm{ISM}}(s)}.
	  \label{eq:veff}
    \end{equation}
    In this work we ignore $\vec{V_{\rm{ISM}}}$.
    We also assume $\psi$ to be 0, which in practice means that we place a lower limit on the screen distance. 

    \subsubsection{The distance of the scattering screens}
    For nine of our pulsars, scintillation arcs can be detected at LOFAR
    frequencies (see Figure~\ref{fig:scintillation_arc_census}). 
    Due to the $\nu^{-2}$ dependence in the arc curvature, scintillation arcs
    at LOFAR frequencies have significantly larger curvature
    and are more diffuse \citep{rszm21}, which complicates their analysis. Arc asymmetries,
    which can provide information on the interplay of dispersion, refraction and
    phase gradients in the IISM \citep{crg+10}, can however be readily quantified.
    %
    
    With equation~\ref{eq:arc_calculate}, we are able to estimate the
    screen distance based on the arc curvature $\eta$ given in
    Table~\ref{tab:arc}.  However, here we clarify that the method
    that we described above has several assumptions in it, namely that the
    screen is isotropic, or that the pulsar
    velocity is aligned with the screen axis ($\psi$ =
    0) and that the screen has no velocity.  The
    influence on distance determination coming from
    $\vec{V_{\rm{ISM}}}$ may be negligible since the screen velocity
    usually has a small value of $\sim$10 km/s compared to the
    transverse velocity of the pulsar for most pulsars, 
    although we note that the screen velocity has in some cases been
    reported to reach (or even exceed) 50 km/s \citep{obv02,rcb+20}.
    The unknown parameter $\cos\psi$ varying between 0 to 1 could
    result in huge discrepancy if the anisotropic screen dominates the
    scattering.  To resolve $\cos\psi$ and $\vec{V_{\rm{ISM}}}$,
    periodic variations in scintillation time-scale or arc curvature,
    or the measure of inter-station time delays is needed
    \citep{rch+19, msa+20, rcb+20, bmg+10}.  In conclusion, under the
    assumptions above, the distance obtained in this work is a
    \emph{lower} limit on the screen distance from Earth, if the screen is
    misaligned with the pulsar velocity.

    \begin{description}
    \item[PSR~J0814+7429:] \cite{rcm00} studied the properties of the
      scattering medium based on long-term weak scintillation
      monitoring.  They proposed that scattering is caused by a
      uniformly extended medium distributed along the entire line of
      sight or located in the range 170--220\,pc from Earth, based on
      a pulsar distance of 310\,pc \citep{tc93}. Since the actual
      distance to this pulsar has since been revised to 433\,pc
      \citep{bbg+02}, we have recomputed the screen distances as
      determined by \citet{rcm00}, which returns a screen in the range
      275--355\,pc.  
      In addition, \cite{bgr98}
      expected the enhanced scattering is located at 72$\pm$13\,pc
      from Earth.  We report for the first time a high degree of asymmetry in the
      arc of this pulsar.  The measured $\eta$ infers a lower limit of
      $D_{s_{low}} = 350\pm13\,$pc on the distance of the
      enhanced scattering material from Earth, consistent with the distanced-revised screen distance from \citet{rcm00}.
    \item[PSR~J0826+2637:] This pulsar has been shown to have four
      arcs \citep{ps06a} at some times and a single arc \citep[e.g.,][]{smc+01}
      at other times.  The arc we observed has $D_{s_{low}} =
      180\pm50\,$pc and is consistent with the "c"
      arc reported by \cite{ps06a} and the single arc with location $240\pm90$\,pc reported by \citet{fab+18}.
    \item[PSR~J0837+0610:] Clear arclets and a 1-ms isolated feature
      were detected and analysed by \cite{bmg+10}. They also reported
      that the screen is located at 420\,pc from Earth which is
      consistent with $460\pm80\,$pc \citep{hsa+05}, 420$\pm$90\,pc
      \citep{fab+18} and 410$\pm$40\,pc \citep{ssa+20} reported earlier. 
      Our result however is
      $D_{\rm s_{low}} = 330\pm30\,$pc from Earth, which is
      only marginally consistent with the other published results. 

    \item[PSR~J0953+0755:] \cite{ssp+14} proposed two enhanced layers
      along the line of sight, at distances of 4.4–16.4\,pc and 26–170\,pc
      to interpret their observations.  We first report an
      asymmetric arc.  From our arc curvature, we
      derive $D_{\rm s_{low}} = 230\pm35\,$pc.

    \item[PSR~J1136+1551:] This pulsar has been detected with four arcs
      \citep{hsb+03,ps06a} and a single arc \citep{sro19} at different times.
      It was also found that a one-dimensional brightness distribution
      is in good agreement with the observed features at
      multiple frequencies with a screen placement of $\ge$136pc from
      Earth \citep{sro19}. This is consistent with our measurement of
      ${D_{\rm s_{low}}} = 120\pm12\,$pc.

    \item[PSR~J1921+2153:] \cite{ssg+17} observed diffractive
      scintillation which they suggested could come from
      inhomogeneities in a thin-screen turbulent plasma at a distance
      of 440\,pc from the observer (based on a pulsar distance of
      1\,kpc from \citealt{cl02}); or from homogeneously distributed scattering material
      between Earth and the pulsar. 
      Re-scaling their screen distance by using the pulsar distance of
      810\,pc derived from the YMW electron-density model
      \citep{ymw17}, their screen distance becomes 360\,pc. This is
      still in disagreement with our value of $D_{\rm s_{low}} =
      500\pm15\,$pc, which we derive from the detection of
      scintillation arcs for this source.   

    \item[PSR~J1932+1059:] This pulsar has been observed with three
      arcs \citep{ps06a} and a single arc \citep{fab+18, yzw+20} at
      different epochs.  The placement of the screen has been
      determined as $200\pm20\,$pc \citep{yzw+20} and $190\pm50\,$pc
      \citep{fab+18}, which is consistent with our result of $D_{\rm
        s_{low}} = 180\pm10\,$pc.

    \item[PSR~J2018+2839:] \cite{fab+18} proposed two possible
      solutions for the distance to this pulsar's scattering screen,
      namely 100\,pc or $550\pm30$\,pc, depending on the analysis
      method used. 
      We found that the scattering screen is likely located at either
      $390\pm45$\,pc or 500$\pm$30\,pc (There are two
      solutions for the measured arc curvature of
      PSR~J2018+2839). The latter of these two solutions
      is highly consistent with the distance \citet{fab+18} derived
      from the scattering time $\tau_{\rm sc}$ and with the angular
      size of the scattering disc as measured by \citet{bgo98}. 

    \item[PSR~J2219+4754:] This pulsar has a highly variable IISM
      along its line of sight \citep{agmk05,mhd+18} and is the first
      pulsar with a clear detection of frequency-dependent, time-variable
      DM \citep{dvt+19}.  The arc curvature of this
      pulsar at 150 MHz is $50.1\pm8.5\,$ s$^{-3}$.  However, this
      curvature does not allow a real distance to be determined, since
      the determinant of Equation~\ref{eq:arc_calculate} becomes
      negative. 
\end{description}
    
    Here, we summarize the properties of the arcs we detected in the
    LOFAR data:
    \begin{enumerate}
      \item Most scintillating pulsars can be seen with arc in our
        census (9/15), which suggests that arcs could be a common
        phenomenon.  Moreover, the highly asymmetry arcs indicate the
        presence of DM gradients, which should be detectable through
        monitoring studies \citep{crg+10}.
    
    \item For a particular pulsar, the number of observable
      scintillation arcs varies with time and observing frequency,
      likely as a consequence of the strong variability of the IISM
      structures involved, or the line-of-sight's rapid motion through
      these structures. Furthermore, the diffuse nature of arcs at low
      frequencies make identification of multiple arcs extremely
      challenging.
    
      \item With the aim of ascertaining the location of the screen,
        knowing the screen orientation $\psi$ could be necessary,
        especially when the screen is highly anisotropic. 
        Once a reliable distance to the screen can be obtained, the types of scattering structures and processes could be confirmed. 
        In our sample,
        PSRs~J0953+0755 and 1932+1059 are surrounded by a nebula
        \citep{rtd+20, mw94, hb08}. Further studies in this direction
        are deferred to a future paper.
\end{enumerate}

    \subsubsection{The spatial scale of the screens}
    We note that the delay axis (Y-axis) of the secondary spectra is
      proportional to the time delay relative to an undeflected ray.
      From this, we can derive the angular extent of the scattering
      material parallel to the direction of the pulsar velocity, as
      given by \citep{crs+06}:
      \begin{equation}
        \theta = \sqrt{\frac{2 \tau s c}{D_{\rm{p}}(1-s)}},
	    \label{eq:theta}
      \end{equation}
      where $\tau$ is time delay derived from the Y-axis of the secondary spectrum (see Figure~\ref{fig:scintillation_arc_census}). This corresponds to a linear extent of 
    \begin{equation}
      L_{s} = 2 D_{\rm p} \left( 1 - s \right) \tan{\theta}.
	  \label{eq:linear_scale}
    \end{equation}
    We find that the spatial scale of the screens we detected are all
    on the order of au (based on the maximum detected delay for our
    scintillation arcs).
    Specifically, the size of the scattering structure for
    PSR~J0837+0610 is 0.12\,au, consistent with the earlier findings
    of 0.2\,au \citep{hsa+05} , but is significantly smaller than the
    16\,au reported by \citet{bmg+10}. The inconsistency between our
    value and that of \citet{bmg+10} is likely due to either lack in
    sensitivity in our LOFAR data (the much higher sensitivity of
    the AO-GBT combination used by \citet{bmg+10} allowed detection of
    arcs out to far greater delays than in our data on this pulsar)
    or to the fact that a different screen dominates the scattering,
    given that there is a larger discrepancy between our scintillation
    bandwidth and theirs. 
    For PSRs~J0953+0755, J1921+2153 and J2018+2839, the interstellar scattering delay is beyond the Nyquist frequency, so only a lower limit on the extent
    of the scattering structure can be given (see Table~\ref{tab:arc})
    and higher frequency resolution is needed to fully determine the
    arc extent.  

    \subsubsection{Impact on Pulsar Timing Arrays}
    In addition to IISM studies discussed in this paper, pulsar
    observations can also be used in high-precision timing experiments
    that have a wide range of applications \citep[see, e.g.][and
      reference therein]{lk05}. Probably the highest-impact such
    experiment is that of the pulsar timing arrays \citep[PTAs,][]{fb90,vob21},
    which aim to use the high rotational stability of radio pulsars to
    detect the faint imprint of extragalactic gravitational waves on
    the space-time metric at Earth. While this experiment has a large
    number of potential noise sources to contend with \citep[see][for
      a review]{vs18}, the time-variable effects of the IISM are
    likely one of the most important ones \citep{lcc+17} and studies
    like ours can help these efforts in a number of ways.

    Currently, PTAs 
    rely on nearby millisecond pulsars to minimize the scattering
    effects, but inclusion of more pulsars would significantly
    increase the sensitivity of the array \citep{sejr13}. Scattering,
    however, reduces the sharpness of the pulse and therefore its
    achievable timing precision; an effect that could be mitigated by
    new methods like cyclic spectroscopy
    \citep{dem11,dsj+21}. Time-variable scattering, however, could
    prove more problematic \citep{hs08,msa+20}. 
 
    Since the observable spatial scale of the scattering screen is
    frequency-dependent, at LOFAR frequencies we can measure changes
    and see features of the scattering screens on much larger angular
    scales, which allows any anomalous scattering features to be
    detected long before they risk contaminating higher-frequency
    observations.  Alternatively, monitoring of refractive effects at
    low frequencies could provide an early-warning system for intense
    IISM studies at higher frequencies, too. The census presented in
    this paper intends to be a first step in this direction, by
    casting some light on the observational requirements for
    high-quality IISM studies at low frequencies, and by identifying
    sources that lend themselves well for such experiments. In
    subsequent papers in this series, we will expand on the results
    shown here by presenting monitoring results and their relation to
    time series of interstellar dispersion, which more directly
    affects PTA timing efforts. 

    \section{Conclusion and future work}

    We have reported the first scintillation census of 31 pulsars with
    LOFAR in the 120-180 MHz frequency band. Large asymmetries in the
    scintillation arcs reflect large-scale gradients of DM. The
    frequency dependencies of $\Delta \nu_{d}$ imply that the
    turbulent features of the interstellar medium deviate from Kolmogorov turbulence at various levels for only a minority of sources (2/15).
    Highly asymmetric arcs from nine pulsars have been
    detected; and these arcs are used to constrain the fractional distance to the phase changing screen.

    Of particular interest are the independent measurement of the
    frequency-scaling factor of $\Delta \nu_{d}$ and the extraordinary measurement of power at an interstellar scattering delay of 3 ms in the secondary spectrum of PSR~J2219+4754 provide meaningful illustrations of the power of low-frequency observations for IISM studies.
    Moreover, low-frequency data have great advantages for echo detection \citep{mhd+18}. Further research into the relations between diffractive scintillation, scintillation arcs and echoes could be very valuable.

    Long term scintillation monitoring with LOFAR of the pulsars
    studied here has commenced and will be reported on in a follow-up
    paper and is useful in testing different scintillation models.  
    Annual variations of scintillation time-scale and arc
    curvature resulting from Earth's motion have been confirmed and
    will be part of that analysis, as well as the study of a possible
    link between DM variations and scintillation arcs.

\begin{acknowledgements}
We thank the anonymous referee for the constructive comments and suggestions, which helped us to improve the presentation of this paper.
The authors thank Dan Stinebring for inspiring discussions that first
got this research started, Bill Coles for useful discussions and
support throughout this work and Vlad Kondratiev for useful discussions and
extensive observation and processing support. 

JPWV acknowledges support by the Deutsche Forschungsgemeinschaft (DFG) through the Heisenberg programme (Project No. 433075039).
YL acknowledges support from the China scholarship council (No. 201808510133).
MB acknowledges funding by the Deutsche Forschungsgemeinschaft (DFG, German Research Foundation) under Germany's Excellence Strategy -- EXC 2121 "Quantum Universe" --  390833306.

Part of this work is based on data obtained with the International LOFAR Telescope (ILT) Core observations under project codes: LC13\_031, LT14\_006, LC15\_030. LOFAR \citep{vwg+13} is the Low Frequency Array designed and constructed by ASTRON. It has observing, data processing, and data storage facilities in several countries, that are owned by various parties (each with their own funding sources), and that are collectively operated by the ILT foundation under a joint scientific policy. The ILT resources have benefitted from the following recent major funding sources: CNRS-INSU, Observatoire de Paris and Universit\'{e} d'Orl\'{e}ans, France; BMBF, MIWF-NRW, MPG, Germany; Science Foundation Ireland (SFI), Department of Business, Enterprise and Innovation (DBEI), Ireland; NWO, The Netherlands; The Science and Technology Facilities Council, UK.

This paper uses and is benefited from data obtained with the German and French LOFAR stations,
during station-owners time and ILT time allocated under project codes
LC0\_014, LC1\_048, LC2\_011, LC3\_029, LC4\_025, LT5\_001, LC9\_039, LT10\_014, LC14\_012, and LC15\_009.
We made use of data from
the Effelsberg (DE601) LOFAR station funded by the Max-Planck-Gesellschaft;
the Unterweilenbach (DE602) LOFAR station funded by
the Max-Planck-Institut für Astrophysik, Garching;
the Tautenburg (DE603) LOFAR station funded by the State of Thuringia,
supported by the European Union (EFRE) and the
Federal Ministry of Education and Research (BMBF) Verbundforschung
project D-LOFAR I (grant 05A08ST1);
the Potsdam (DE604) LOFAR station funded by the
Leibniz-Institut für Astrophysik, Potsdam;
the Jülich (DE605) LOFAR station supported by the
BMBF Verbundforschung project DLOFAR I (grant 05A08LJ1);
and the Norderstedt (DE609) LOFAR station funded by the
BMBF Verbundforschung project D-LOFAR II (grant 05A11LJ1).
The observations of the German LOFAR stations
were carried out in stand-alone GLOW mode,
which is technically operated and supported by
the Max-Planck-Institut für Radioastronomie, the Forschungszentrum
Jülich and Bielefeld University. We acknowledge support and
operation of the GLOW network, computing and storage facilities by
the FZ-Jülich, the MPIfR and Bielefeld University and financial support
from BMBF D-LOFAR III (grant 05A14PBA) and D-LOFAR IV (grant 05A17PBA),
and by the states of Nordrhein-Westfalia and Hamburg.
We acknowledge the work of A.~Horneffer in setting up
the GLOW network and initial recording machines.
LOFAR station FR606 is hosted by the Nan\c{c}ay Radio Observatory and
is operated by Paris Observatory, associated with the French Centre
National de la Recherche Scientifique (CNRS) and Universit\'{e} d'Orl\'{e}ans.
\end{acknowledgements}

\bibliographystyle{aa}
\bibliography{./journals,./psrrefs,./modrefs,./crossrefs}

    \appendix
    
    \section{Scintillation Bandwidth from Literature}
    \label{appendix:a}
    The table below summarises all previously published values for the
    scintillation bandwidth of the pulsars included in our study. The
    values given are also included in
    Figure~\ref{fig:scintillation_frequency_rescaling}.\\
    
    \topcaption{Previously published measurements of scintillation bandwidth $\Delta
      \nu_{\rm{d}}$ for pulsars included in this paper.}
    \label{Table:Scintillation_Bandwidth_from_Literature}
    \tablefirsthead
    {\hline \hline PSR &\multicolumn{1}{c}{Freq.} & \multicolumn{1}{c}{$\Delta\nu_{\rm{d}}$} & \multicolumn{1}{c}{Ref.} \\ 
      Name &\multicolumn{1}{c}{(MHz)} & \multicolumn{1}{c}{(MHz)} & \multicolumn{1}{c}{} \\ \hline}
    \tablehead{%
      \multicolumn{4}{c}%
                  {{\bfseries  Continued from previous column}} \\
                  \hline \hline PSRs J&\multicolumn{1}{c}{Freq.} & \multicolumn{1}{c}{$\Delta\nu_{\rm{d}}$} & \multicolumn{1}{c}{Ref.} \\ 
                  \multicolumn{1}{c}{}  &\multicolumn{1}{c}{(MHz)} & \multicolumn{1}{c}{(MHz)} & \multicolumn{1}{c}{} \\ \hline}
    \tabletail{%
      \midrule \multicolumn{4}{r}{{Continued on next column}} \\ \midrule}
    \tablelasttail{%
      \\\midrule
      \bottomrule}
      
    \begin{supertabular}{lccr}
     J0034$-$0721 & 327 & 1.0(2)          & 1 \\
     (B0031$-$07) & 408 & 4               & 2 \\ 
                & 436 & 3.80              & 3\\
                & 660 & 7.55              & 3\\
     J0332+5434 &111  & 0.000243          & 4 \\
     (B0329+54) & 327 & 0.130(4); 0.165(13); 60 & 5, 1, 6  \\
                & 340 & 0.036             & 7\\
                & 408 & 0.03; 0.047; 0.088; 0.8              & 2, 8, 7, 9\\
                & 610 & 4.0; 130; 220; 349               & 9, 6, 10\\
                &1000 & 3.89              & 11\\
                & 1420& 5930              & 6\\
                & 1540& 9.2(2.2); 14(1)   & 12, 13\\
                & 2250& 17(2); 20(2); 67(14)          & 14\\
     J0814+7429 & 41 & 0.0020(6)          & 15\\
     (B0809+74) &51.5 & 0.0039(7)         & 16\\
                & 62.43 & 0.007(3)        & 15\\
                &81.5 & 0.023(1)          & 16\\
                & 88.57 & 0.02(1)         & 15\\
                & 111.87& 0.045(5)        & 15\\
                & 151 & 1.2               & 9\\
                & 408 & 4; $\geq30$       & 2, 9\\
                &1000 & 85.114            & 11\\
     J0826+2637 &102.7& 0.0014$\pm$0.0003 & 17\\
     (B0823+26) & 326.5 & 0.058           & 18\\
                & 327 & 0.24(2); 0.250(8); & \\
                &     & 0.29(2); 0.29(4) & 5, 1 \\
                & 408 & 0.189; 0.5        & 8, 2\\
                & 430 & 0.215; 0.396      & 19\\
                &1000 & 33.113            & 11\\
                & 1540& 82(5)             & 13\\
                & 1700& 81(3)             & 20 \\
     J0837+0610 &102.7& 0.0023(2)         & 17\\
     (B0834+06) & 324 & 0.0040(5); 0.19(1); 0.21(1); & \\
                &     & 0.24(2); 0.28(2); 0.35(2)  & 21\\
                & 326.5 & 0.495           & 18\\
                & 327 & 0.353(8); 0.37(1);  & \\
                &     & 0.42(1); 0.45(3);  &  \\
                &     & 0.49(2); 0.62(2); 0.003 & 5, 1, 22 \\
                & 335 & 0.350             & 23\\
                & 408 & 1.260; 1.6; 5.5   & 8, 2, 9\\
                & 410 & 0.280; 1.45       & 2, 23\\
                & 430 & 0.05; 0.136; 0.151; 0.223; & \\
                &     & 0.243; 0.396; 0.406             & 19\\
                &1000 & 25.704            & 11\\
      J0953+0755 & 41 & 0.0015(4)         & 15\\
      (B0950+08) & 47 & 0.028(2)          & 24 \\
                & 51 & 0.019(1); 0.031(4); 0.049(7)  & 24\\
                & 62.43 & 0.03(1)         & 15\\
                & 88.57 & 0.10(4)         & 15\\
                & 111.87& 0.22(6)         & 15 \\
                & 154& 4.1                & 25\\
                & 320 & 0.792; 1.188      & 19\\
                & 327 & $\geq$9           & 1\\
                & 340 & $\geq$1.44        & 7\\
                & 408 & $\geq$2.65; $\geq20$        & 7, 9\\
                & 430 & 1.089             & 19\\
                & 436 & 2.00              & 3\\
                & 450 & $\geq$1.75        & 7\\
                &1000 & 162.181           & 11\\
     J1136+1551 &102.7& 0.0082(6)         & 17\\
     (B1133+16) & 326.5 & 0.710           & 18\\
                & 327 & 0.43(2); 0.82(6); &   \\
                &     & 0.84(2); 1.40(3)  & 5, 1 \\
                & 340 & 0.590             & 7\\
                & 408 & $\geq$0.97; 1.3; $\geq10$        & 7, 2, 9\\
                & 430 & 0.743; 0.782;     &  \\
                &     & 0.990; 1.155      & 19\\
                & 450 & $\geq$1.52        & 7\\
                &1000 & 60.260            & 11\\
                &1540 & 134(11); 149(20); & \\
                &     & 163(38)    & 26\\
     J1239+2453 &102.7& 0.009(1)          & 17\\
     (B1237+25) & 326.5 & 0.595           & 18\\
                & 327 & 1.15(6); 1.8(1)   & 5, 1 \\
                & 340 & $\geq$0.67        & 7\\ 
                & 408 & $\geq$0.99; $\geq4$        & 7, 2\\
                & 430 & 0.693; 0.842      & 19\\
                &1000 & 61.660            & 11\\
                &1540 & 127(22); 154(18)  & \\
                &     & ; 156(25)    & 26\\
     J1607$-$0032 & 320 & 0.248; 0.376    & 19\\
     (B1604$-$00) & 326.5 & 0.165         & 18\\
                & 327 & 0.376(15); 0.38(2) & 5, 1\\
                & 335 & 0.020             & 23\\
                & 408 & $\geq$4           & 2\\
                & 410 & 0.035; 0.05       & 23 \\
                & 430 & 1.089             & 19\\
                & 630 & 0.670             & 23\\
                &1000 & 51.286            & 11\\
     J1921+2153 &102.5& 8.6               & 4 \\
     (B1919+21) &102.7& 0.0055(7)         & 17\\
                & 320 & 0.0074; 0.019;    &   \\
                &     & 0.02; 0.025; 0.054            & 19\\
                & 324 & 0.33              & 27\\
                & 326.5 & 0.330           & 18\\
                & 327 & 0.269(9); 0.29(1); &  \\
                &     & 0.30(2); 0.55(1)  & 5, 1 \\
                & 335 & 0.730             & 23\\
                & 408 &  0.9; 3           & 2, 9\\
                & 410 & 2.400             & 23\\
                & 430 & 0.099; 0.105; 0.119; & \\
                &     &  0.129; 0.149; 0.347             & 19\\
                &1000 & 23.988            & 11\\
     J1932+1059 & 320 & 0.614; 0.743; 0.891; 1.19             & 19\\
     (B1929+10) & 327 & 1.20(8); 1.29(3); 60 & 5, 1, 6 \\
                & 335 & 1.600             & 23\\
                & 408 & 2.2               & 2\\
                & 410 & 5.7; 1.0          & 23\\
                &1000 & 138.038           & 11\\
                & 1540& 268(24)           & 13\\
     J2018+2839 &102.7& 0.0016(2)         & 17\\
     (B2016+28) & 326.5 & 0.132           & 3\\
                & 327 & 0.201(8); 0.21(1) & 5, 1 \\
                & 408 & 0.092; 0.3        & 8, 2\\
                & 430 & 0.094; 0.104; 0.129; 0.14; 0.149; 0.151             & 19\\
                & 610 & 0.63              & 6\\
                &1000 & 6.310             & 11\\
     J2022+2854 & 320 & 0.064             & 19\\
     (B2020+28) & 326.5 & 0.396           & 18\\
                & 327 & 0.199(5); 0.27(2) & 5, 1 \\
                & 408 & 0.56; 0.6         & 7, 2\\  
                & 430 & 0.129; 0.148; 0.158; 0.178; 0.183             & 19\\
                & 450 & $\geq$0.83        & 7\\
                &1000 & 13.490            & 11\\
                & 1540& 70(5)             & 13\\
     J2022+5154 & 408 & 0.4               & 2\\
     (B2021+51) & 610 & 0.81; 3.41        & 6\\
                &1000 & 2.291             & 11\\
                & 1540& 52(3)             & 13\\
     J2219+4754 & 408 & 0.056             & 7\\
     (B2217+47) &1000 & 2.29              & 11 \\
    \end{supertabular}
        \\{\bf Notes:}  Given are the pulsar name, the observing frequency, the reported scintillation bandwidth and the corresponding literature reference given as superscripted letters are listed below the table.
    \begin{enumerate}
    \item[$^{\rm{ref}}$] References: (1) \citealt{bgr98}, (2) \citealt{sw85}, (3) \citealt{jnk98}, (4) \citealt{kps+01}, (5) \citealt{brg99}, (6) \citealt{spg+17}, (7) \citealt{ar81}, (8) \citealt{grl94}, (9) \citealt{r70}, (10) \citealt{sfm96}, (11) \citealt{c86}, (12) \citealt{wym+08}, (13) \citealt{wmj+05}, (14) \citealt{whh+18}, (15) \citealt{ss08}, (16) \citealt{bgt+21}, (17) \citealt{mss+95}, (18) \citealt{bk85}, (19) \citealt{cwb85}, (20) \citealt{dlk13}, (21) \citealt{ssa+20}, (22) \citealt{bmg+10}, (23) \citealt{ra82}, (24) \citealt{pc92}, (25) \citealt{bmj+16}, (26) \citealt{new13}, (27) \citealt{ssg+17}, (28) This work.
    \end{enumerate}

\end{document}